# Optimal Operation Method for Computing Power-Electric Power Coordination Considering End-to-End Completion Latency of Computing Tasks

LIU Yize[1], YAN Mingyu[1]*, WEN Jianfeng[2], SONG Meng[3], YANG Qing[4], Mariusz Malinowski[5]

(1. State Key Laboratory of Advanced Electromagnetic Engineering and Technology (School of Electrical and Electronic Engineering, Huazhong University of Science and Technology), Wuhan 430074, Hubei Province, China;2. Department of Electrical Engineering and Electronics, University of Liverpool, Liverpool L69 3GJ, U.K.;3. School of Electrical Engineering, Southeast University, Nanjing 210096, Jiangsu Province, China;4. State Key Laboratory of Coal Combustion, Huazhong University of Science and Technology, Wuhan 430074, Hubei Province, China;5. Warsaw University of Technology, Warsaw 00-661, Poland)

**ABSTRACT:** With the rapid growth of computing demand and the large-scale integration of renewable energy, how to realize the coordinated optimal operation of computing networks and power networks has become an important issue to be addressed. However, existing studies mainly focus on the impacts of spatiotemporal migration of computing loads on power system operation, while the internal task processing procedures of computing networks are largely neglected. To address this issue, this paper proposes an optimal operation method for computing power–electric power coordination considering the end-to-end completion latency of computing tasks. First, a full-process latency model covering the “transmission–buffering–computation” procedure of computing tasks is established, which uniformly characterizes forwarding waiting, network transmission, queueing, and computation processing. Second, a task-level spatiotemporal scheduling mechanism is developed to jointly optimize the task forwarding time, routing path, and destination computing node. Then, a computing power–electric power coordinated optimization model is formulated to minimize the power supply cost of the power network and the total completion latency of computing tasks. Case studies demonstrate that the proposed method can fully exploit the task-level spatiotemporal scheduling flexibility of computing loads and facilitate the spatiotemporal matching between computing loads and renewable energy, thereby reducing the system power supply cost while ensuring the quality of service for computing tasks.

## 0 Introduction

The rapid development of emerging technologies, including large language models, autonomous driving, virtual reality, and digital twins, is continuously increasing the scale and electricity demand of computing infrastructure [1]. According to the International Energy Agency, global data-center electricity consumption reached approximately 415 TWh in 2024, accounting for 1.5% of global electricity use, and is expected to rise to 945 TWh by 2030 [2]. Meanwhile, installed wind and photovoltaic capacity continues to expand. The simultaneous growth of energy-intensive computing loads and high shares of renewable energy creates new challenges for power-system balancing and economical operation [3]. China’s 15th Five-Year Plan therefore calls for coordinated deployment of green electricity and computing power [4]. Unlike conventional inflexible loads, computing tasks can adjust their processing time and location while meeting quality-of-service requirements, thereby changing the spatiotemporal distribution of computing loads [5,6]. Exploiting this flexibility to match computing loads with renewable generation can improve renewable-energy utilization and reduce system operating costs.

Many studies have investigated computing-load models and flexibility. Energy-consumption models have evolved from equipment-level descriptions of servers and cooling systems [7-9] to task-level models for general computing tasks [10] and large-language-model training [11]. Regarding temporal flexibility, studies [12,13] use the deferrable nature of delay-tolerant tasks to shift computing loads while satisfying completion deadlines and quality-of-service requirements. Regarding spatial flexibility, study [14] introduces a calibratable probabilistic latency-bound constraint to guarantee service quality under interregional data-center scheduling. Studies [15,16] examine dynamic coalition formation and benefit allocation among multiple data centers. Considering uncertainty in computing demand and renewable generation, studies [17-19] optimize interregional task migration in the "Eastern Data, Western Computing" setting. Study [20] develops a coordinated computing-power optimization method based on edge-fog-cloud task offloading and heterogeneous hierarchical data centers.

However, these studies mainly describe how spatiotemporal migration of computing loads affects power-grid operation and largely neglect the internal task-processing procedure of the computing network. In practice, task migration depends not only on the processing time and destination computing node, but also on network topology, link capacity, node-buffer states, and computing resources [21]. Existing studies lack a unified model of the complete transmission-buffering-computation process and therefore cannot jointly optimize forwarding time, routing path, and destination node under computing-network constraints. Consequently, they cannot accurately quantify end-to-end completion latency or fully exploit the spatiotemporal scheduling potential of computing loads.

To address these limitations, this paper proposes a coordinated computing power－electric power operation method that considers the end-to-end completion latency of computing tasks. First, a full-process latency model decomposes task completion into forwarding wait, network transmission, period synchronization, queueing, and computation processing. Second, a task-level spatiotemporal scheduling mechanism changes processing time through forwarding-time decisions and changes load location through joint routing and destination-node decisions. Finally, a coordinated optimization model balances power-network economy and computing-network service timeliness. The power network optimizes generator output and power flow according to computing-load demand, while the computing network optimizes the spatiotemporal distribution of tasks, enabling coordinated matching between computing loads and renewable generation while satisfying task service-quality requirements.

## 1 Architecture and Operation Mechanism of the Coordinated System

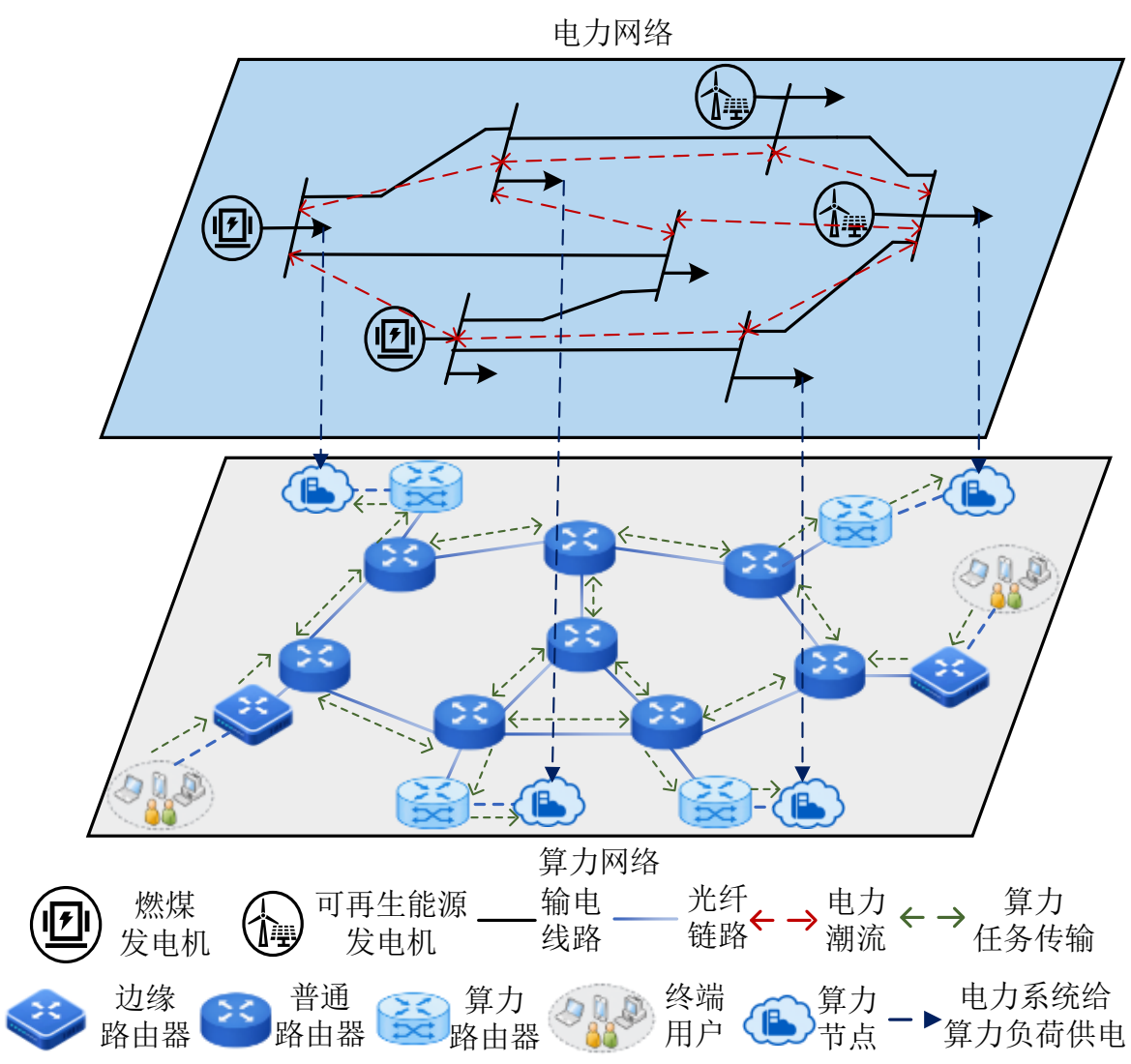

**Fig. 1 Architecture of the coordinated electricity-computing power systems**

The proposed coordinated system is shown in Fig. 1. It consists of a power network and a computing network. Coal-fired units and renewable generation supply the computing network, while the computing network provides task access, transmission, and computation. After end users generate tasks with different computing requirements and strict latency limits, the tasks enter through edge routers, select suitable forwarding times, pass through ordinary routers and optical-fiber links to computing routers, and finally reach the corresponding computing nodes for processing.

Computing loads provide both temporal and spatial scheduling flexibility. Temporally, changing a task's forwarding time changes its processing time and shifts the computing load. Spatially, optimizing the routing path and destination computing node changes the processing location and shifts the computing load among regions.

Using this flexibility, the coordinated system jointly considers user demand, computing capacity, communication resources, renewable generation, coal-fired generation, and power-flow information. It jointly optimizes task forwarding periods, routing paths, and computing nodes while balancing power-network economy and computing-network service timeliness. This coordination matches computing loads with renewable output in both time and space, improves renewable-energy utilization, reduces power supply cost, and satisfies task deadlines.

## 2 Coordinated Optimization Model

### 2.1 Objective Function

To balance power-network operating economy and computing-network service timeliness, the objective minimizes the sum of the power-network supply cost and the total completion latency of computing tasks, as follows:(1)

$$\min \sum_{t\in\mathcal{T}} \sum_{g\in\mathcal{G}} C_{g,t} + \lambda \sum_{k\in\mathcal{K}} D_k^{\text{sum}} \tag{1}$$

where the sets correspond to scheduling periods, coal-fired units, and computing tasks, respectively; the operating cost is defined for coal-fired unit g in period t; the task term is the total completion latency of task k; and the final parameter is the weighting coefficient for total task completion latency. $\mathcal{T}$ $\mathcal{G}$ $\mathcal{K}$ $C_{g,t}$ $D_k^{\text{sum}}$ $\lambda$

### 2.2 Computing-Network Model

This subsection establishes the mathematical model of the computing network [21]. After a task enters through an edge router, its scheduling depends on its source, data size, computing demand, and completion deadline. Task k is represented by the following attribute vector:

$$\omega_k = \left(i_k^{\text{src}}, d_k, c_k, T_k^{\text{arr}}, \bar{T}_k\right) \tag{2}$$

where the vector describes computing task k; the router term identifies the edge router through which task k enters the network; the data and computation terms denote its data size in bits and required processor clock cycles, respectively; and the time terms denote its arrival time and latest completion time. $\omega_k$ $i_k^{\text{src}}$ $d_k$ $c_k$ $T_k^{\text{arr}}$ $\bar{T}_k$

#### 2.2.1 Task-Level Spatiotemporal Scheduling Decisions

Spatiotemporal scheduling assigns each computing task a unique computing node, routing path, and forwarding time, as follows:(3)(5)

$$\sum_{m\in\mathcal{M}} x_{k,m} = 1,\ \forall k \tag{3}$$

$$\sum_{p\in\mathcal{P}_k} y_{k,p} = 1,\ \forall k \tag{4}$$

$$\sum_{t\in\mathcal{T}} z_{k,t} = 1,\ \forall k \tag{5}$$

where the sets contain computing nodes and candidate routing paths. The node, path, and forwarding-time binary variables equal 1 when task k selects computing node m, routing path p, and forwarding period t, respectively, and equal 0 otherwise. $\mathcal{M}$ $\mathcal{P}_k$ $x_{k,m}$ $y_{k,p}$ $z_{k,t}$

Because the endpoint of the selected routing path must be consistent with the selected computing node, the node and path decisions satisfy:

$$\sum_{p\in\mathcal{P}_{k,m}} y_{k,p} = x_{k,m},\ \forall k, \forall m \tag{6}$$

where the set contains the candidate paths from the edge router of task k to computing node m. $\mathcal{P}_{k,m}$

According to the forwarding-period decision, the forwarding start time of task k is:

$$T_k^s = \sum_{t\in\mathcal{T}} (t-1)\Delta t z_{k,t},\ \forall k \tag{7}$$

where the first term is the forwarding start time of task k and the second parameter is the duration of one scheduling period. $T_k^s$ $\Delta t$

A computing task cannot be forwarded before it arrives at the edge router:

$$T_k^s \ge T_k^{\text{arr}},\ \forall k \tag{8}$$

#### 2.2.2 Data-Network Transmission Model

A computing task is transmitted link by link along the selected path to its destination node. Transmission latency consists of data-transmission latency and signal-propagation latency. The former depends on task data size and link transmission rate, whereas the latter depends on physical link length and propagation speed. The network transmission latency of task k along path p is expressed as:

$$D_{k,p,t}^{\text{tr}} = \sum_{e\in\mathcal{E}_p} \left(\frac{d_k}{r_{e,t}} + \frac{L_e}{v}\right),\ \forall k, \forall p, \forall t \tag{9}$$

where the latency term represents transmission of task k through path p in period t; the path set contains its communication links; the rate term is the data-transmission rate of link l in period t; the length term is the length of link e; and v is signal propagation speed in the medium. $D_{k,p,t}^{\text{tr}}$ $\mathcal{E}_p$ $r_{e,t}$ $L_e$

#### 2.2.3 Node Buffer-Queue Model

Each computing node buffers and processes tasks according to first-in, first-out service. After arriving at its destination, a task enters the buffer queue and is executed in arrival order. The queue changes dynamically as new tasks arrive and existing tasks are processed, and its state transition is:

$$Q_{m,t+1} = Q_{m,t} + A_{m,t} - W_{m,t},\ \forall m, \forall t \tag{10}$$

$$A_{m,t} = \sum_{k \in \mathcal{K}} c_k \delta_{k,m,t},\ \forall m, \forall t \tag{11}$$

$$0 \le Q_{m,t} \le Q_m^{\max},\ \forall m, \forall t \tag{12}$$

where the terms denote the unprocessed workload, arriving workload, and completed workload of computing node m in period t. The task-arrival variable equals 1 when task k arrives at node m in period t and equals 0 otherwise. The final parameter is the node's buffer-capacity limit. $Q_{m,t}$ $A_{m,t}$ $W_{m,t}$ $\delta_{k,m,t}$ $Q_m^{\max}$

2.2.4 Task Computation Model

Because computing resources are limited, the workload completed by a node in one period cannot exceed either the pending workload or the node's maximum processing capacity:(13)

$$W_{m,t} = \min\{Q_{m,t} + A_{m,t}, f_m \Delta t\},\ \forall m, \forall t \tag{13}$$

where the parameter is the computing rate of node m, expressed as executable processor clock cycles per unit time. $f_m$

To ensure that all tasks received during the scheduling horizon are completed, the computing buffer at the beginning and end of the horizon must satisfy:

$$Q_{m,1} = 0,\ \forall m \tag{14}$$

$$Q_{m,\mathrm{T}+1} = 0,\ \forall m \tag{15}$$

where the two terms are the unprocessed workloads at computing node m at the beginning and end of the scheduling horizon, respectively. $Q_{m,1}$ $Q_{m,\mathrm{T}+1}$

2.2.5 Full-Process Task-Latency Model

From arrival at the edge router to completion, a computing task undergoes forwarding wait, network transmission, period synchronization, queueing, and computation processing.

After arriving at the edge router, a task waits until its selected forwarding time. Its forwarding-wait latency is:

$$D_k^{\mathrm{fw}} = T_k^{\mathrm{s}} - T_k^{\mathrm{arr}},\ \forall k \tag{16}$$

where the variable is the forwarding-wait latency of computing task k. $D_k^{\mathrm{fw}}$

Network transmission latency depends on the routing path and forwarding time. Under the scheduling decisions, the actual transmission latency of task k is:

$$D_k^{\mathrm{tr}} = \sum_{t \in \mathcal{T}} \sum_{p \in \mathcal{P}_k} y_{k,p} z_{k,t} D_{k,p,t}^{\mathrm{tr}},\ \forall k \tag{17}$$

where the variable is the actual transmission latency of computing task k. $D_k^{\mathrm{tr}}$

Under discrete-time scheduling, a task that completes network transmission must wait until the start of the next period before processing. This produces period-synchronization latency, as follows:(18)(19)

$$T_k^{\mathrm{a}} = \left\lceil \frac{T_k^{\mathrm{s}} + D_k^{\mathrm{tr}}}{\Delta t} \right\rceil \Delta t,\ \forall k \tag{18}$$

$$D_k^{\mathrm{al}} = T_k^{\mathrm{a}} - T_k^{\mathrm{s}} - D_k^{\mathrm{tr}},\ \forall k \tag{19}$$

where the terms denote the time at which task k enters the computing queue, its period-synchronization latency, and the ceiling operator, respectively. $T_k^{\mathrm{a}}$ $D_k^{\mathrm{al}}$ $\lceil \cdot \rceil$

After a task enters the computing buffer, its queueing latency consists of the unfinished workload already present and the workload of other tasks arriving in the same period. Previously arrived tasks are processed first-in, first-out. Because the order of tasks arriving within the same period is uncertain, those tasks are processed in random order. The queueing latency of task k arriving at node m in period t is:

$$D_{k,m,t}^{\mathrm{q}} = \frac{Q_{m,t}}{f_m} + \frac{A_{m,t} - c_k}{2 f_m},\ \forall k, \forall m, \forall t \tag{20}$$

where the variable denotes the queueing latency after task k arrives at node m in period t. $D_{k,m,t}^{\mathrm{q}}$

The computation-processing latency depends on task workload and the computing resources of the selected node:(21)

$$D_{k,m}^{\mathrm{c}} = \frac{c_k}{f_m},\ \forall k, \forall m \tag{21}$$

Given the computing-node, forwarding-time, and routing-path decisions, the total completion latency and completion time of task k are:

$$D_k^{\mathrm{sum}} = D_k^{\mathrm{fw}} + D_k^{\mathrm{tr}} + D_k^{\mathrm{al}} + \sum_{m \in \mathcal{M}} \sum_{t \in \mathcal{T}} \delta_{k,m,t} D_{k,m,t}^{\mathrm{q}} + \sum_{m \in \mathcal{M}} x_{k,m} D_{k,m}^{\mathrm{c}},\ \forall k \tag{22}$$

$$T_k^{\mathrm{f}} = T_k^{\mathrm{arr}} + D_k^{\mathrm{sum}},\ \forall k \tag{23}$$

where the terms denote the total completion latency and completion time of task k, and the set contains all computing nodes. $D_k^{\mathrm{sum}}$ $T_k^{\mathrm{f}}$ $\mathcal{M}$

To meet the quality-of-service requirement, the completion time must not exceed the task deadline:

$$T_k^{\mathrm{f}} \le \overline{T}_k\ \forall k \tag{24}$$

2.2.6 Energy Consumption and Supply Model of the Computing Network

Computing-network energy consumption is mainly caused by task transmission and computation processing.

The computing network uses an IP-over-fiber routing architecture. IP routers switch and forward data in the electrical domain, while optical signals carry data over fiber links. Task data therefore requires electronic processing at routers and optical-electrical/electrical-optical conversion by transponders. The associated electronic-processing and signal-conversion power is calculated as follows:(25)(26)

$$P_{k,p,t}^{\mathrm{tr}} = \frac{d_k}{\Delta t}\left[\phi_i^{\mathrm{IP}} + \sum_{j\in\mathcal{R}_p}\phi_j^{\mathrm{IP}} + 2\left(\phi_i^{\mathrm{op}} + \phi_m^{\mathrm{op}} + \sum_{j\in\mathcal{R}_p}\phi_j^{\mathrm{op}}\right)\right]$$

$, \forall k, \forall p, \forall t$

(25)

$$P_k^{\mathrm{tr}} = \sum_{t\in\mathcal{T}}\sum_{p\in\mathcal{P}_k} y_{k,p} z_{k,t} P_{k,p,t}^{\mathrm{tr}},\ \forall k \tag{26}$$

where the first terms denote the transmission power of task k over path p in period t and under the actual schedule; the path-related set contains ordinary routers on path p; the router parameters denote unit-data electronic-processing energy at edge router i and ordinary router j; and the transponder parameters denote unit-data conversion energy at edge router i, ordinary router j, and computing router m. $P_{k,p,t}^{\mathrm{tr}}$ $P_k^{\mathrm{tr}}$ $\mathcal{R}_p$ $\phi_i^{\mathrm{IP}}$ $\phi_j^{\mathrm{IP}}$ $\phi_i^{\mathrm{op}}$ $\phi_j^{\mathrm{op}}$ $\phi_m^{\mathrm{op}}$

Computing nodes consume energy when executing tasks. According to processor dynamic-power characteristics, the computation power of node m in period t depends on its computing rate and completed workload:

$$P_{m,t}^{\mathrm{IT}} = \frac{\zeta_m f_m^2 W_{m,t}}{\Delta t},\ \forall m, \forall t \tag{27}$$

where the terms denote the computation power of node m in period t and the equivalent capacitance parameter of its computing chip. $P_{m,t}^{\mathrm{IT}}$ $\zeta_m$

Each computing node is supplied jointly by local renewable generation and the power network:(28)(30)

$$P_{m,t}^{\mathrm{out}} = P_{m,t}^{\mathrm{IT}} - P_{m,t}^{\mathrm{R}},\ \forall m, \forall t \tag{28}$$

$$0 \le P_{m,t}^{\mathrm{R}} \le \bar{P}_{m,t}^{\mathrm{R}},\ \forall m, \forall t \tag{29}$$

$$0 \le P_{m,t}^{\mathrm{out}},\ \forall m, \forall t \tag{30}$$

where the terms denote power supplied by the grid to computing node m in period t, locally consumed renewable power, and the renewable-generation upper limit in the region containing node m. $P_{m,t}^{\mathrm{out}}$ $P_{m,t}^{\mathrm{R}}$ $\bar{P}_{m,t}^{\mathrm{R}}$

2.3 Power-Network Model

This subsection establishes the mathematical model of the power network [22].

The output-power constraint of each coal-fired unit is:

$$0 \le P_{g,t}^{\mathrm{G}} \le P_g^{\max},\ \forall g, \forall t \tag{31}$$

where the terms denote the output of coal-fired unit g in period t and its maximum output. $P_{g,t}^{\mathrm{G}}$ $P_g^{\max}$

The transmission-line capacity constraint is:

$$-F_l^{\max} \le P_{l,t}^{\mathrm{L}} \le F_l^{\max},\ \forall l, \forall t \tag{32}$$

where the terms denote the power flow of line l in period t and its maximum capacity. $P_{l,t}^{\mathrm{L}}$ $F_l^{\max}$

The DC power-flow constraint is:

$$P_{l,t}^{\mathrm{L}} \cdot X_l = S_{\mathrm{base}}\left(\theta_{l_{\mathrm{from}},t} - \theta_{l_{\mathrm{to}},t}\right),\ \forall l, \forall t \tag{33}$$

where the terms denote the reactance of line l, the system base power, and the voltage phase angles at the sending and receiving buses of line l in period t. $X_l$ $S_{\mathrm{base}}$ $\theta_{l_{\mathrm{from}},t}$ $\theta_{l_{\mathrm{to}},t}$

The nodal power-balance constraint is:

$$\sum_{l\in\mathcal{L}} K_{b,l}^{\mathrm{L}} P_{l,t}^{\mathrm{L}} = \sum_{g\in\mathcal{G}} K_{b,g}^{\mathrm{P}} P_{g,t}^{\mathrm{G}} - P_{b,t}^{\mathrm{D}},\ \forall b, \forall t \tag{34}$$

where the sets contain transmission lines and coal-fired units; the load term is the demand at bus b in period t; and the matrices describe bus-line and bus-generator incidence, respectively. $\mathcal{L}$ $\mathcal{G}$ $P_{b,t}^{\mathrm{D}}$ $K_{b,l}^{\mathrm{L}}$ $K_{b,g}^{\mathrm{P}}$

The nodal voltage-angle constraints are:

$$-\pi \le \theta_{b,t} \le \pi,\ \forall b, \forall t \tag{35}$$

$$\theta_{b_{\mathrm{ref}},t} = 0,\ \forall t \tag{36}$$

where the terms denote the voltage phase angle of bus b and that of the reference bus in period t. $\theta_{b,t}$ $\theta_{b_{\mathrm{ref}},t}$

The operating cost of a coal-fired unit is represented by a quadratic function:

$$C_{g,t} = \alpha_{0,g} + \alpha_{1,g} P_{g,t}^{\mathrm{G}} + \alpha_{2,g}\left(P_{g,t}^{\mathrm{G}}\right)^2,\ \forall g, \forall t \tag{37}$$

where the cost term is the operating cost of coal-fired unit g in period t, and the remaining parameters are coefficients of its operating-cost function. $C_{g,t}$ $\alpha_{0,g}$ $\alpha_{1,g}$ $\alpha_{2,g}$

2.4 Coupling Constraint

The computing and power networks are coupled through the electrical loads of the computing nodes:(38)

$$P_{b_m,t}^{D} = P_{m,t}^{\mathrm{out}},\ \forall m, \forall t \tag{38}$$

where the parameter identifies the power-system bus connected to computing node m. $b_m$

# 3 Case Studies

The proposed method is evaluated on two test systems: a 6-bus/7-node coordinated system and a 118-bus/47-node coordinated system. Simulations are performed in MATLAB R2024a on a computer with an Intel Core i9-14900HX CPU at 2.20 GHz and 32 GB of memory, using Gurobi 13.0.2.

3.1 6-Bus/7-Node Coordinated System

The topology is shown in Fig. 2. The computing network contains one edge router, two ordinary routers, two computing routers, and two computing nodes. The power network contains six buses, eleven transmission lines, three coal-fired units, and one photovoltaic generator. The photovoltaic generator is directly connected to computing node S1 and its output is consumed locally. The computing capacities of S1 and S2 are 1800 and 3600 Gcycle/s, respectively, and the transmission rates of links A-B, A-C, and B-C vary over time.

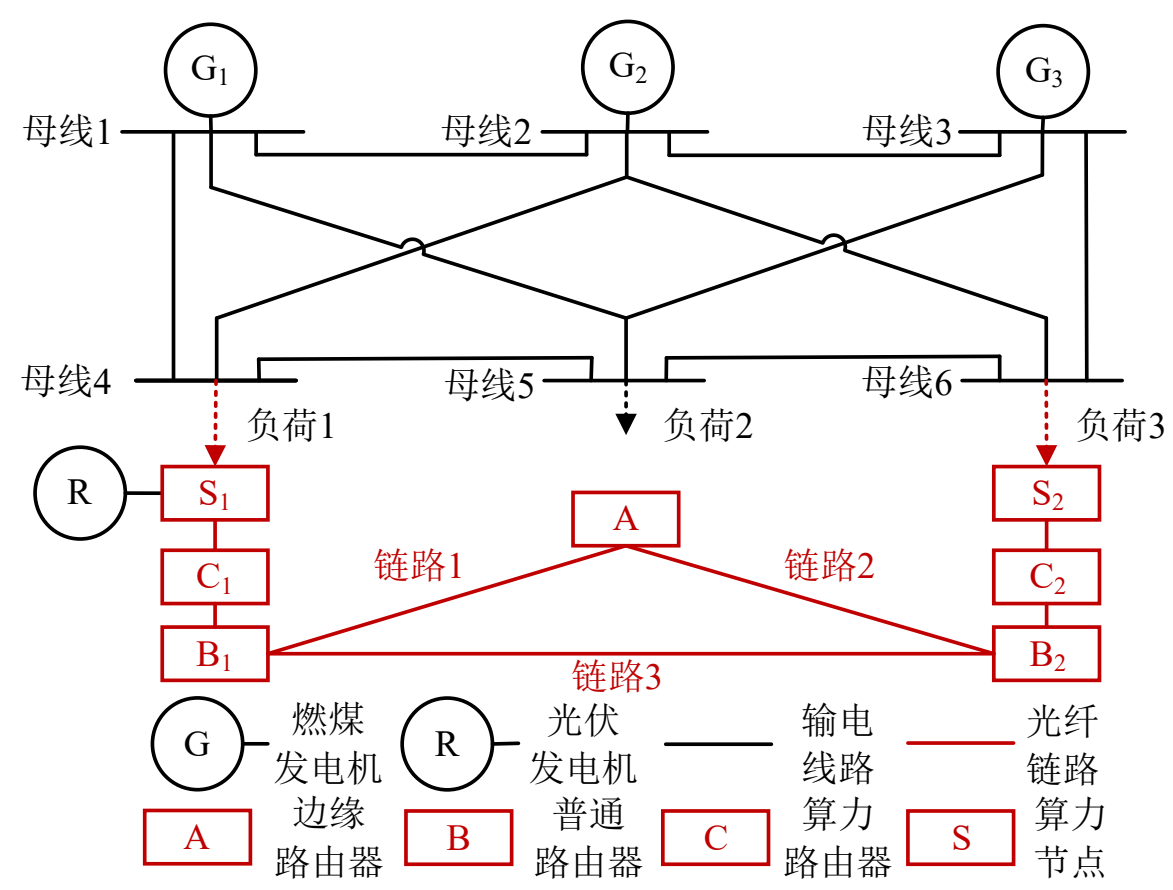

**Fig. 2 Topology of the 6-7 node power-computing coordinated system**

**Tab. 1 Basic parameters of computing tasks**

| Task | Arrival time | Deadline | Data size/GB | Workload/Pcycle |
|---|---|---|---|---|
| 1 | 0：00 | 4：00 | 20.00 | 32.40 |
| 2 | 1：00 | 17：00 | 35.00 | 36.00 |
| 3 | 6：00 | 12：00 | 22.00 | 9.60 |
| 4 | 12：00 | 19：00 | 26.00 | 19.20 |
| 5 | 1：00 | 7：00 | 25.00 | 38.40 |

To illustrate the transmission-buffering-computation process, five tasks with different data sizes and computing demands are configured in the 6-bus/7-node system, as listed in Table 1.

**Tab. 2 Energy consumption and supply composition of computing nodes**

| Computing node | Transmission energy/kWh | Computation energy/kWh | Renewable supply/kWh | 燃煤发电机供电/kWh |
|---|---|---|---|---|
| S1 | 0.012 | 360.000 | 278.330 | 81.670 |
| S2 | 0.006 | 393.330 | 0 | 393.330 |
| Total | 0.018 | 753.330 | 278.330 | 475.000 |

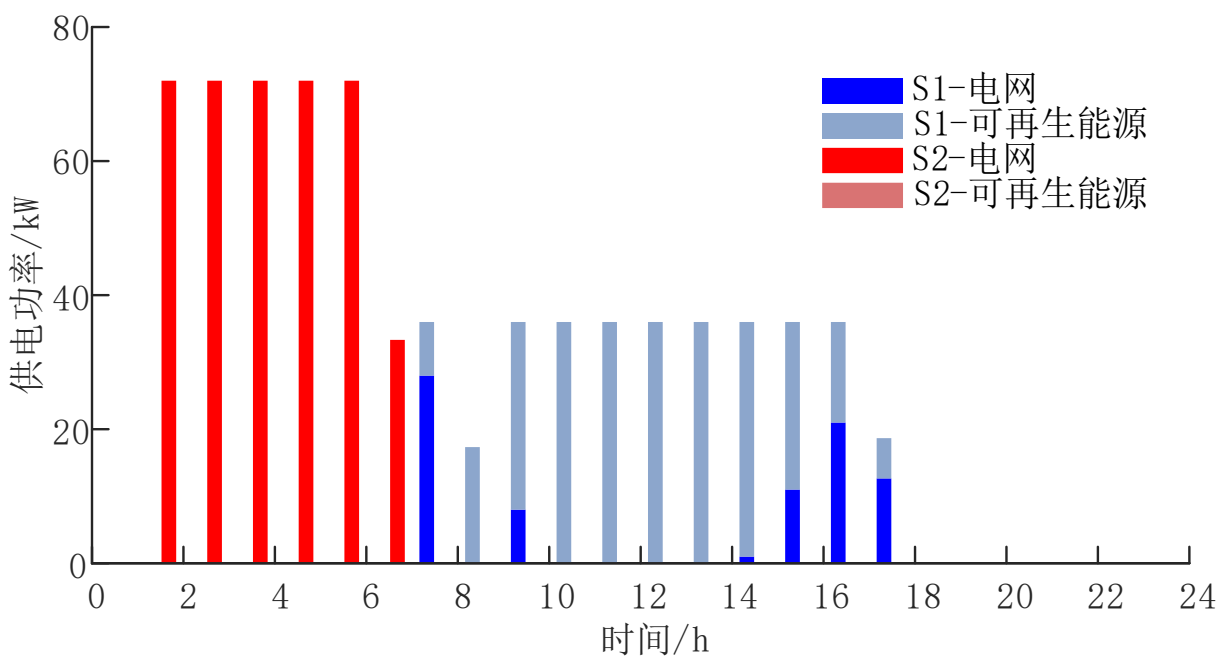

**Fig. 3 Hourly power supply of computing nodes**

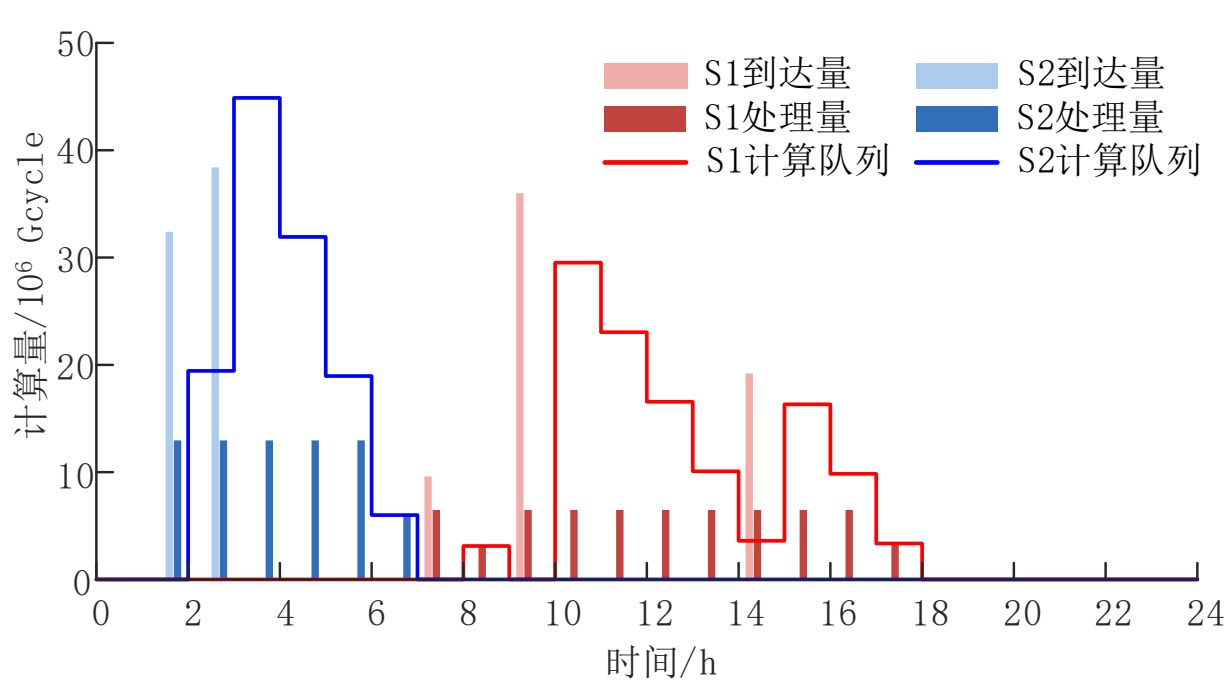

**Fig. 4 Task arrivals and computational processing at computing nodes**

**Tab. 3 Spatiotemporal Dispatch Results of Tasks in the Computing Network**

| Task | Computing node | Path | Forwarding time |
|---|---|---|---|
| 1 | S2 | A-C-E-S2 | 0：00 |
| 2 | S1 | A-B-D-S1 | 8：00 |
| 3 | S1 | A-C-B-D-S1 | 6：00 |
| 4 | S1 | A-B-D-S1 | 13：00 |
| 5 | S2 | A-C-E-S2 | 1：00 |

Regarding routing, Tasks 1, 2, 4, and 5 select paths containing fewer links, whereas Task 3 selects A-C-B-D-S1. Its alternatives are A-B-D-S1 and A-C-B-D-S1. From 06:00 to 07:00, the transmission rates of links A-B, A-C, and C-B are 40, 145, and 130 Mbit/s, respectively. Although A-C-B-D-S1 contains more links, its total transmission latency is lower under these time-varying rates.

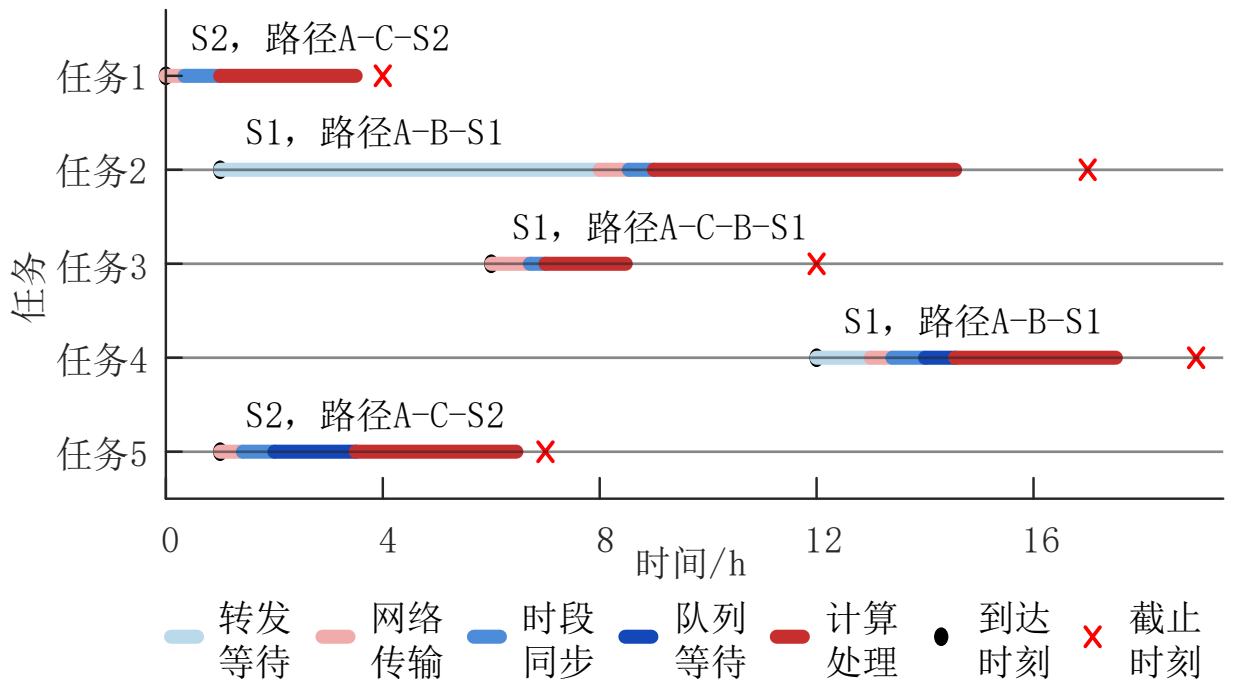

**Fig. 5 Overall processing timeline of computing tasks**

### 3.2 118-Bus/47-Node Coordinated System

The computing network contains five edge routers, twenty-two ordinary routers, ten computing routers, and ten computing nodes, based on the European GÉANT topology [23]. The power network uses the IEEE 118-bus system with 186 transmission lines, 54 coal-fired units, four photovoltaic generators, and three wind generators. Renewable generators are assigned according to regional resource availability: nodes 1, 4, 7, and 10 are connected to photovoltaic generation, while nodes 2, 5, and 8 are connected to wind generation.

Four scenarios are defined to verify the proposed method:

Scenario I: baseline operation without task flexibility; tasks follow the initial schedule.

Scenario II: temporal flexibility only; task forwarding times are optimized.

Scenario III: spatial flexibility only; computing nodes and routing paths are optimized.

Scenario IV: both temporal and spatial flexibility; forwarding times, computing nodes, and routing paths are jointly optimized.

**Tab. 4 Comparison of system operational performance under different scenarios**

| Scenario | Renewable supply/MWh | Grid supply/MWh | Renewable utilization/% | Power supply cost/CNY | Reduction/% |
|---|---|---|---|---|---|
| I | 206.70 | 1321.88 | 21.19 | 55145.59 | 0 |
| II | 414.41 | 1114.16 | 42.49 | 50846.25 | 7.80 |
| III | 827.95 | 700.62 | 84.90 | 42379.68 | 23.15 |
| IV | 945.10 | 583.47 | 96.91 | 39984.21 | 27.49 |

| Scenario | Forwarding-wait latency/h | Transmission latency/h | Synchronization latency/h | Queueing latency/h | Processing latency/h | Average completion latency/h |
|---|---|---|---|---|---|---|
| I | 0.00 | 0.28 | 0.72 | 0.67 | 0.17 | 1.84 |
| II | 1.03 | 0.28 | 0.72 | 0.38 | 0.17 | 2.58 |
| III | 0.00 | 0.45 | 0.57 | 1.75 | 0.20 | 2.97 |
| IV | 1.23 | 0.42 | 0.58 | 0.06 | 0.18 | 2.47 |

Scenario II uses temporal migration only. With computing nodes and paths fixed, task forwarding times are adjusted to match renewable output. Relative to Scenario I, the total cost decreases by 7.80% and renewable-energy utilization rises to 42.49%. However, some tasks are delayed, increasing forwarding-wait latency to 1.03 h and average completion latency to 2.58 h.

Scenario III uses spatial migration only. Tasks are forwarded immediately, but their computing nodes and paths are optimized according to regional renewable output. Some tasks migrate to renewable-rich nodes, reducing total cost by 23.15% and increasing renewable-energy utilization to 84.90%. However, workload concentration causes queue congestion. Queueing latency reaches 1.75 h, the highest among the four scenarios, and average completion latency increases to 2.97 h.

Scenario IV jointly uses temporal and spatial migration. Tasks are processed during periods of high renewable output and distributed among regions according to renewable availability, while queue congestion is avoided. Relative to Scenario I, total cost decreases by 27.49% and renewable-energy utilization rises to 96.91%. Forwarding-wait latency increases to 1.23 h because some tasks are delayed, but spatial optimization reduces queueing latency to 0.06 h and average completion latency to 2.47 h.

Overall, the proposed method jointly optimizes task forwarding time, routing path, and destination node. It improves renewable-energy utilization, coordinates computing-resource allocation, and balances power-network economy with computing-network service quality.

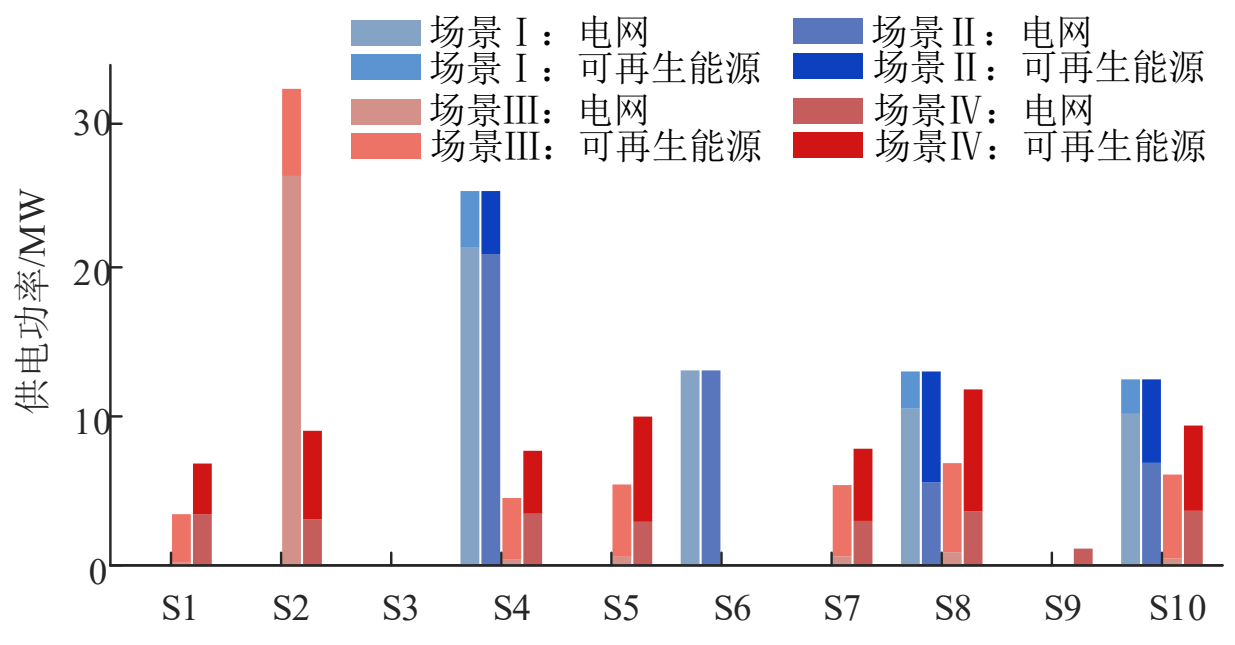

**Fig. 6 Hourly power supply of computing nodes under different scenarios**

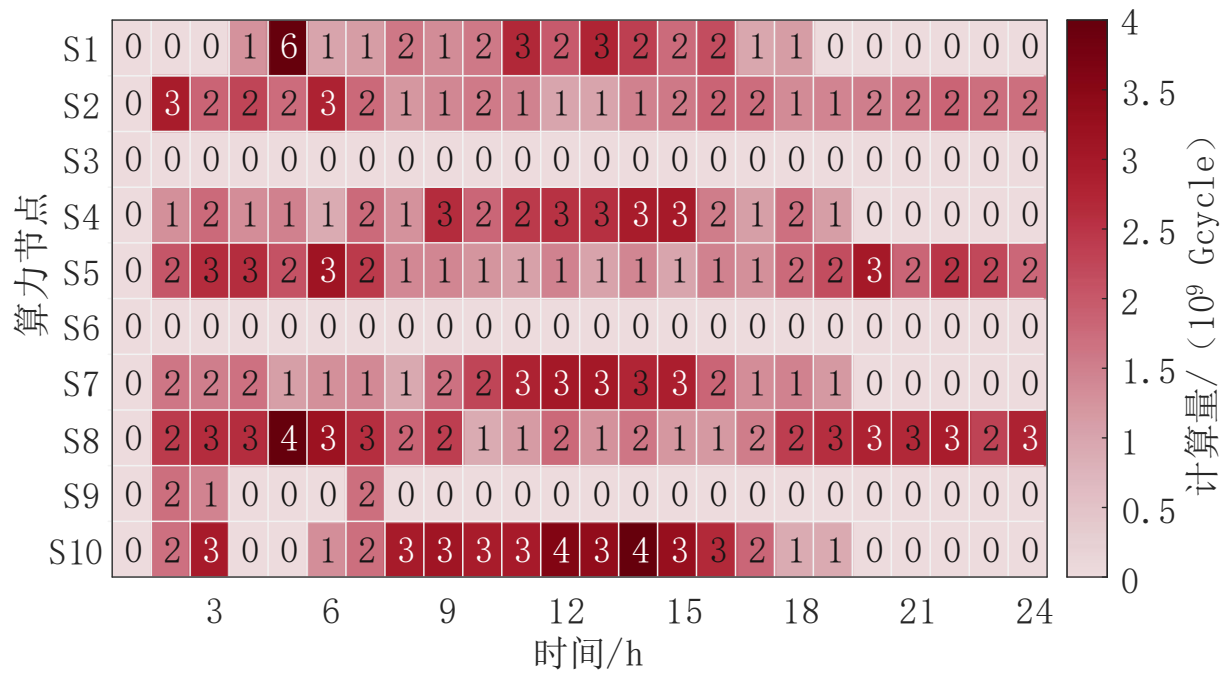

**Fig. 7 Hourly processed computing workload of computing nodes in Scenario IV**

## 4 Conclusion

This paper proposes a coordinated computing power - electric power operation method that considers end-to-end task completion latency and matches computing loads with renewable generation in time and space. First, it uniformly models the complete transmission-buffering-computation process and quantifies end-to-end latency. Second, it jointly optimizes forwarding periods, routing paths, and destination nodes to provide task-level spatiotemporal flexibility. Third, it formulates a coordinated optimization model that balances power-network economy and computing-network service timeliness. Case studies demonstrate that the method reduces power supply cost while maintaining task service quality. Future work will consider uncertainty in task arrivals and renewable generation to improve scheduling robustness.

# 计及算力任务端到端完成时延的电算协同优化运行方法

刘祎泽[1]，晏鸣宇[1]*，文剑峰[2]，宋梦[3]，杨晴[4]，Mariusz Malinowski[5]

（1.强电磁工程与新技术国家重点实验室(华中科技大学电气与电子工程学院),湖北省 武汉市 430074；
2．利物浦大学电气工程与电子系，英国 利物浦 L69 3GJ；
3．东南大学电气工程学院，江苏省 南京市 210096；
4．煤燃烧国家重点实验室（华中科技大学），湖北省 武汉市 430074；
5．华沙理工大学，波兰 华沙 00-661）

## Optimal Operation Method for Computing Power-Electric Power Coordination Considering End-to-End Completion Latency of Computing Tasks

LIU Yize[1], YAN Mingyu[1]*, WEN Jianfeng[2], SONG Meng[3], YANG Qing[4], Mariusz Malinowski[5]

(1. State Key Laboratory of Advanced Electromagnetic Engineering and Technology (School of Electrical and Electronic Engineering, Huazhong University of Science and Technology), Wuhan 430074, Hubei Province, China;2. Department of Electrical Engineering and Electronics, University of Liverpool, Liverpool L69 3GJ, U.K.;3. School of Electrical Engineering, Southeast University, Nanjing 210096, Jiangsu Province, China;4. State Key Laboratory of Coal Combustion, Huazhong University of Science and Technology, Wuhan 430074, Hubei Province, China;5. Warsaw University of Technology, Warsaw 00-661, Poland)

**ABSTRACT:** With the rapid growth of computing demand and the large-scale integration of renewable energy, how to realize the coordinated optimal operation of computing networks and power networks has become an important issue to be addressed. However, existing studies mainly focus on the impacts of spatiotemporal migration of computing loads on power system operation, while the internal task processing procedures of computing networks are largely neglected. To address this issue, this paper proposes an optimal operation method for computing power – electric power coordination considering the end-to-end completion latency of computing tasks. First, a full-process latency model covering the “transmission – buffering – computation” procedure of computing tasks is established, which uniformly characterizes forwarding waiting, network transmission, queueing, and computation processing. Second, a task-level spatiotemporal scheduling mechanism is developed to jointly optimize the task forwarding time, routing path, and destination computing node. Then, a computing power – electric power coordinated optimization model is formulated to minimize the power supply cost of the power network and the total completion latency of computing tasks. Case studies demonstrate that the proposed method can fully exploit the task-level spatiotemporal scheduling flexibility of computing loads and facilitate the spatiotemporal matching between computing loads and renewable energy, thereby reducing the system power supply cost while ensuring the quality of service for computing tasks.

**摘要：**随着算力需求快速增长和可再生能源大规模接入，如何实现算力网络与电力网络的协同优化运行，已成为亟待解决的重要问题。然而，现有研究主要侧重于描述算力负荷时空迁移对电网运行的影响，忽略了算力网络内部任务处理过程。为此，本文提出一种计及算力任务端到端完成时延的电算协同优化运行方法。首先，建立了算力网络内算力任务“传输-缓存-计算”全流程时延模型，统一描述转发等待、网络传输、队列等待、计算处理等环节。其次，建立了任务级时空调度机制，联合优化任务的转发时间、路由路径和目标算力节点。然后，以电力网络供电成本和算力网络的算力任务总完成时延最小为优化目标，构建了电算协同优化模型。算例结果表明，所提方法可以充分挖掘算力负荷任务级时空调度潜力，引导算力负荷与可再生能源在时空上协调匹配，从而降低系统供电成本并保证算力任务服务质量。

## 0 引言

随着大语言模型、自动驾驶、虚拟现实和数字孪生等新兴技术的快速发展，推动算力基础设施规模及用电需求持续增长[1]。国际能源署统计表明，2024 年全球数据中心用电量约 415 TW·h，占全球用电量的 1.5%，预计 2030 年将增至 945 TW·h[2]。与此同时，以风电、光伏为代表的可再生能源装机规模不断提升。高耗能算力负荷持续增长与高比例可再生能源接入对电力系统供需平衡和经济运行提出了新的挑战[3]。为应对上述挑战，国家“十五五”规划纲要明确提出推动绿色电力与算力协同布局[4]。与传统刚性负荷相比，算力任务可在满足服务质量要求的条件下调整处理时刻和位置，从而改变算力负荷的时空分布[5,6]。因此，充分挖掘算力负荷的时空调节潜力，引导其与可再生能源出力在时空上协调匹配，实现算力网络与电力网络的协同运行，对提升可再生能源消纳水平、降低系统运行成本具有重要意义。

目前，已有很多针对算力负荷建模及其灵活性的研究。在算力负荷能耗建模方面，现有研究由服务器、冷却系统等数据中心设备能耗建模[7-9]，逐步向通算任务[10]和大语言模型训练负荷[11]等任务级精细化建模拓展。在算力负荷时间灵活性方面，文献[12,13]基于延迟容忍性任务的可延迟执行特性，在满足任务完成时限和服务质量要求的前提下实现算力负荷的时间迁移与灵活调节。在算力负荷空间灵活性方面，文献[14]通过引入可校准的概率时延上界约束，保障了跨区域数据中心空间负荷灵活调度下的服务质量。文献[15,16]从构建多数据中心联盟的角度，研究了联盟动态匹配构建及其内部利益协调分配机制。文献[17-19]考虑计算需求和可再生能源出力不确定性，面向东数西算场景，通过跨区域任务迁移优化算力负荷分布，促进可再生能源消纳。文献[20]基于边-雾-云任务卸载，构建了计及分层数据中心异构特性的算力-电力联合优化调度方法。

然而，上述研究主要侧重于描述算力负荷的时空迁移能力对电网运行的影响，忽略了算力网络内部任务处理过程。实际上，算力任务的时空迁移不仅取决于任务处理时刻和目标算力节点，还受到算力网络拓扑、链路传输能力、节点缓存状态及计算资源的共同制约[21]。现有研究尚缺乏对算力网络内部任务“传输-缓存-计算”全过程的统一建模，无法在算力网络运行约束下联合优化任务转发时刻、路由路径与目标算力节点，进而难以准确量化任务端到端完成时延从而保障任务服务质量，也难以充分挖掘算力负荷的时空调度潜力。

基于以上分析，本文提出了一种计及算力任务端到端完成时延的电算协同优化运行方法。首先，构建算力任务“传输-缓存-计算”全流程时延模型，将任务完成过程分解为转发等待、网络传输、时段同步、队列等待和计算处理环节，实现任务端到端完成时延的统一刻画；然后，提出任务级时空调度机制，通过调整任务的转发时间改变计算处理时间，联合优化路由路径和目标算力节点改变算力负荷的空间分布，挖掘算力负荷的时空调度潜力；最后构建兼顾电力网络运行经济性与算力网络的服务时效性的电算协同优化模型，电力网络侧根据算力负荷需求优化机组出力和网络潮流，算力网络通过任务调度优化算力负荷的时空分布，实现算力负荷与可再生能源出力在时空维度上协调匹配，从而降低系统供电成本并保证算力任务服务质量。

## 1 电算协同系统架构与运行机制

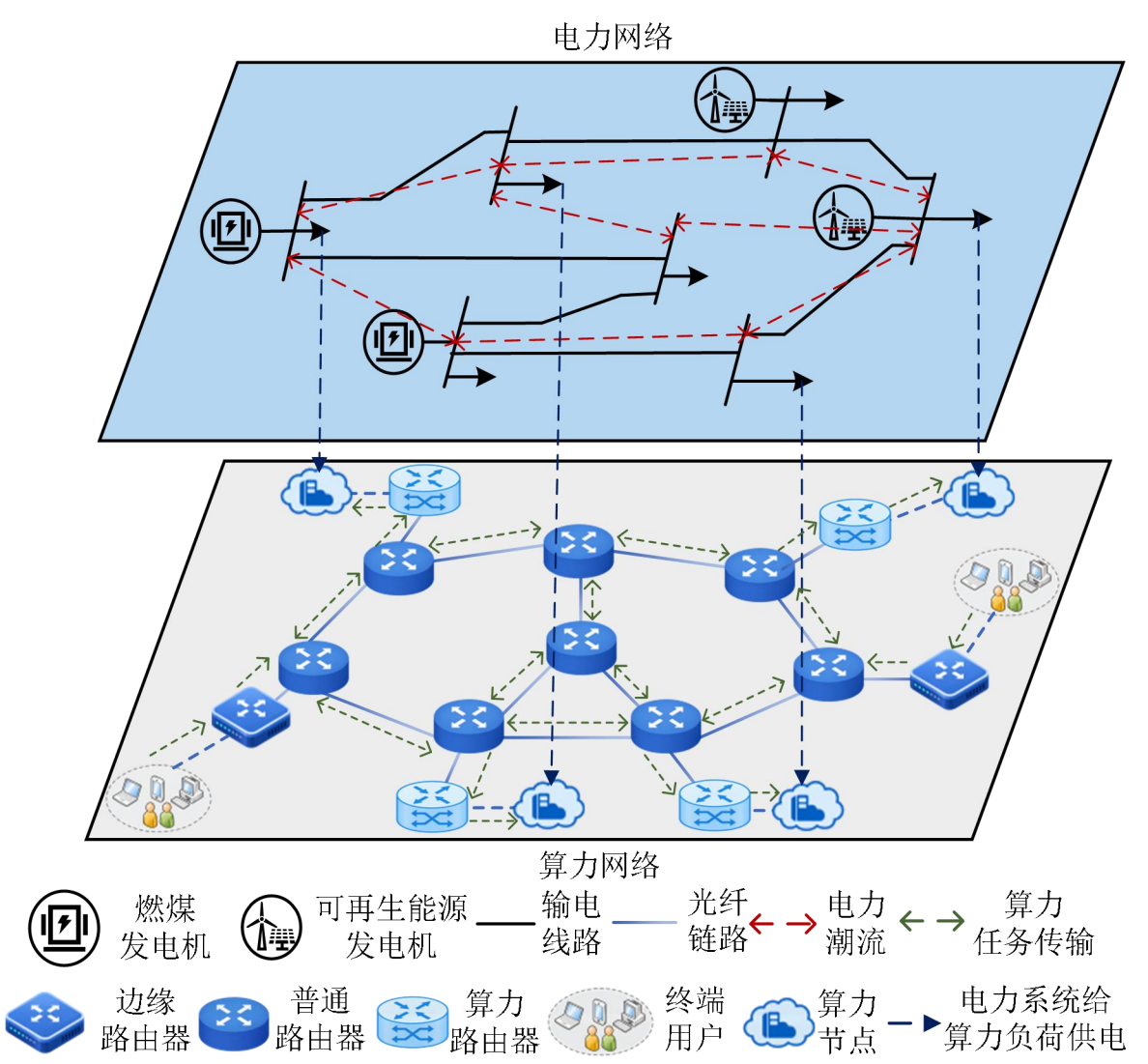

**图 1 电算协同系统架构**

**Fig. 1 Architecture of the coordinated electricity-computing power systems**

本文构建的电算协同系统架构如图 1 所示。该系统由电力网络和算力网络构成，电力网络通

过燃煤机组和可再生能源为算力网络供能；算力网络负责算力任务的接入、传输与计算，终端用户生成具有不同计算需求和严格时延要求的算力任务后，任务首先经边缘路由器接入算力网络，选择合适的转发时间，经普通路由器与光纤链路传输至算力路由器，然后到达对应的算力节点完成计算。

算力负荷具有时空调度灵活性，在时间维度上，可通过调整任务的转发时间改变其计算处理时刻，实现算力负荷的时间迁移；在空间维度上，通过优化路由路径和目标算力节点，改变任务的计算位置，实现算力负荷的空间迁移。

基于上述时空调节能力，电算协同系统综合考虑算力网络的用户需求、计算能力和通信资源以及电力网络的可再生能源出力、燃煤机组出力及网络潮流等运行信息，兼顾电力网络运行经济性与算力网络的服务时效性，联合优化算力任务转发时段、路由路径和算力节点。在满足算力任务完成时限等服务质量要求的同时，引导算力负荷与可再生能源出力在时空维度上协调匹配，从而提高可再生能源消纳水平并降低系统供电成本。

## 2 电算协同优化运行模型

### 2.1 目标函数

为兼顾电力网络运行经济性与算力网络的服务时效性，以电力网络供电成本和算力网络的算力任务总完成时延最小为优化目标，如式(1)所示：

$$\min \sum_{t\in\mathcal{T}}\sum_{g\in\mathcal{G}} C_{g,t} + \lambda \sum_{k\in\mathcal{K}} D_k^{\text{sum}} \tag{1}$$

式中，$\mathcal{T}$ 、$\mathcal{G}$ 和 $\mathcal{K}$ 分别为调度时段、燃煤机组和算力任务集合；$C_{g,t}$ 为燃煤机组 $g$ 在时段 $t$ 的运行费用；$D_k^{\text{sum}}$ 是算力任务 $k$ 的总完成时延；$\lambda$ 为算力任务总完成时延项的权重系数。

### 2.2 算力网络模型

本小节建立算力网络的数学模型[21]。算力任务经边缘路由器接入算力网络后，其调度过程受到任务来源、数据规模、计算需求及完成时限等属性影响。为描述不同算力任务的基本特征，将任务 $k$ 可表示为：

$$\omega_k = \left(i_k^{\text{src}}, d_k, c_k, T_k^{\text{arr}}, \overline{T}_k\right) \tag{2}$$

式中，$\omega_k$ 为算力任务 k 的属性向量；$i_k^{\text{src}}$ 为任务 $k$ 接入算力网络的边缘路由器；$d_k$ 和 $c_k$ 分别为任务 $k$ 的数据量（以比特数表示）和计算量（执行该任务所需处理器的时钟周期数）；$T_k^{\text{arr}}$ 为任务 $k$ 到达边缘路由器的时刻；$\overline{T}_k$ 为任务 $k$ 最晚完成时刻。

#### 2.2.1 任务时空调度决策

算力任务的时空调度为各算力任务确定唯一的算力节点、路由路径和转发时间，如式(3)-(5)所示：

$$\sum_{m\in\mathcal{M}} x_{k,m} = 1,\ \forall k \tag{3}$$

$$\sum_{p\in\mathcal{P}_k} y_{k,p} = 1,\ \forall k \tag{4}$$

$$\sum_{t\in\mathcal{T}} z_{k,t} = 1,\ \forall k \tag{5}$$

式中，$\mathcal{M}$ 为算力节点集合；$\mathcal{P}_k$ 为算力任务的可选路由路径集合；$x_{k,m}$ 为算力节点决策变量，当任务$k$选择算力节点$m$时为 1,否则为 0；$y_{k,p}$ 为路由路径决策变量，当任务 $k$ 选择路由路径 $p$ 时为 1,否则为 0；$z_{k,t}$ 为转发时间决策变量，当任务 $k$ 选择在时段 $t$ 开始转发时为 1，否则为 0。

由于所选路由路径的终点应与所选算力节点保持一致，节点与路径之间满足：

$$\sum_{p\in\mathcal{P}_{k,m}} y_{k,p} = x_{k,m},\ \forall k,\forall m \tag{6}$$

式中，$\mathcal{P}_{k,m}$ 为算力任务 $k$ 从其边缘路由器到算力节点 $m$ 的可选路径集合。

根据任务的转发时段决策结果，算力任务 $k$ 的转发开始时刻为：

$$T_k^s = \sum_{t\in\mathcal{T}} (t-1)\Delta t z_{k,t},\ \forall k \tag{7}$$

式中，$T_k^s$ 为任务 $k$ 的转发开始时刻；$\Delta t$ 为单个调度时段的时长。

算力任务不能在到达边缘路由器之前进行转发：

$$T_k^{\text{s}} \ge T_k^{\text{arr}},\ \forall k \tag{8}$$

#### 2.2.2 数据网络传输模型

算力任务经选定的路由路径逐条传输至目标算力节点，其任务传输时延主要由数据发送时延和信号传播时延两部分构成。数据发送时延是指将任务数据完整发送至通信链路所需的时间，取决于任务数据量和链路传输速率，信号传播时延是指信号在通信介质中传播链路距离所需时间，取决于链路的物理长度和信号传播速度。任务 $k$ 经路径 $p$ 传输时的网络传输时延表示为：

$$D_{k,p,t}^{\text{tr}} = \sum_{e\in\mathcal{E}_p}\left(\frac{d_k}{r_{e,t}} + \frac{L_e}{v}\right),\ \forall k,\forall p,\forall t \tag{9}$$

式中，$D_{k,p,t}^{\text{tr}}$ 为任务 $k$ 在时段 $t$ 经路径 $p$ 的传输时延；$\mathcal{E}_p$ 为路径 $p$ 所包含的通信链路集合；$r_{e,t}$ 为链路 $l$ 在时段 $t$ 的数据传输速率；$L_e$ 为链路 $e$ 的长度；$v$ 为信号在传输介质中的传播速度。

#### 2.2.3 节点缓存队列模型

每个算力节点遵循先进先出原则对算力任务

进行缓存与处理，算力任务到达目标算力节点后进入任务缓存队列，并按照到达顺序依次执行。随着新任务到达及任务处理过程，算力节点的任务缓存队列动态变化，其状态转移关系为：

$$Q_{m,t+1}=Q_{m,t}+A_{m,t}-W_{m,t}\,,\ \forall m,\forall t \tag{10}$$

$$A_{m,t}=\sum_{k\in\mathcal{K}}c_k\delta_{k,m,t}\,,\ \forall m,\forall t \tag{11}$$

$$0\le Q_{m,t}\le Q_m^{\max},\ \forall m,\forall t \tag{12}$$

式中，$Q_{m,t}$ 为算力节点 $m$ 在时段 $t$ 尚未执行的计算量；$A_{m,t}$ 为算力节点 $m$ 在时段 $t$ 到达的计算量；$W_{m,t}$ 是算力节点 $m$ 在时段 $t$ 完成的计算量；$\delta_{k,m,t}$ 为任务到达状态变量，当任务 $k$ 在时段 $t$ 到达算力节点 $m$ 时为 1，其他为 0；$Q_m^{\max}$ 为算力节点 $m$ 任务缓存队列容量上限。

2.2.4 任务计算处理模型

受算力节点计算资源的限制，节点在单个时段内完成的计算量不能超过当前待处理任务量和节点的最大可处理计算量，如式(13)所示：

$$W_{m,t}=\min\{Q_{m,t}+A_{m,t},f_m\Delta t\},\ \forall m,\forall t \tag{13}$$

式中，$f_m$ 为算力节点 $m$ 的计算速率，以单位时间内可执行的处理器时钟周期数表示。

为保证调度周期内接收到的算力任务全部处理完成，算力节点在调度周期开始和结束时的计算缓存队列应满足：

$$Q_{m,1}=0,\ \forall m \tag{14}$$

$$Q_{m,\mathrm{T}+1}=0,\ \forall m \tag{15}$$

式中，$Q_{m,1}$ 和 $Q_{m,\mathrm{T}+1}$ 分别为算力节点 $m$ 在调度周期开始和结束时尚未执行的计算量。

2.2.5 任务全流程时延模型

算力任务从到达边缘路由器至计算完成依次经历转发等待、网络传输、时段同步，队列等待和计算处理过程。

算力任务到达边缘路由器后，需等待至对应的转发时刻再进行转发，其从到达至开始转发所经历的转发等待时延为：

$$D_k^{\mathrm{fw}}=T_k^{\mathrm{s}}-T_k^{\mathrm{arr}}\,,\ \forall k \tag{16}$$

式中，$D_k^{\mathrm{fw}}$ 为算力任务 $k$ 的转发等待时延。

网络传输时延取决于路由路径和转发时刻，根据相应的调度决策，算力任务 $k$ 的实际传输时延为：

$$D_k^{\mathrm{tr}}=\sum_{t\in\mathcal{T}}\sum_{p\in\mathcal{P}_k}y_{k,p}z_{k,t}D_{k,p,t}^{\mathrm{tr}}\,,\ \forall k \tag{17}$$

式中，$D_k^{\mathrm{tr}}$ 为算力任务 $k$ 的实际传输时延。

由于离散时间调度机制，任务完成网络传输后，需等待至下一时段起始时刻进行处理，所以会产生时段同步时延，如式(18)和(19)所示：

$$T_k^{\mathrm{a}}=\left\lceil\frac{T_k^{\mathrm{s}}+D_k^{\mathrm{tr}}}{\Delta t}\right\rceil\Delta t,\ \forall k \tag{18}$$

$$D_k^{\mathrm{al}}=T_k^{\mathrm{a}}-T_k^{\mathrm{s}}-D_k^{\mathrm{tr}},\ \forall k \tag{19}$$

式中，$T_k^{\mathrm{a}}$ 为算力任务 $k$ 进入算力节点计算队列的时刻；$D_k^{\mathrm{al}}$ 为算力任务 $k$ 的时段同步时延；$\lceil . \rceil$ 为向上取整符号。

算力任务进入计算缓存队列后，其队列等待时延由两部分构成：任务到达前算力节点中尚未完成的任务引起的等待时延和与该任务在同一时段到达的其他任务所引起的等待时延。对于此前到达且尚未完成的任务，算力节点遵循先进先出原则依次处理；对于同一时段到达的任务，由于其到达顺序的不确定性，算力节点遵循随机执行原则。算力任务 $k$ 在时段 $t$ 到达算力节点 $m$ 的队列等待时延表示为：

$$D_{k,m,t}^{\mathrm{q}}=\frac{Q_{m,t}}{f_m}+\frac{A_{m,t}-c_k}{2f_m},\ \forall k,\forall m,\forall t \tag{20}$$

式中，$D_{k,m,t}^{\mathrm{q}}$ 为任务 $k$ 在时段 $t$ 到达算力节点 $m$ 后的队列等待时延。

算力任务的计算处理时延取决于任务的计算量和算力节点的计算资源，如式(21)所示：

$$D_{k,m}^{\mathrm{c}}=\frac{c_k}{f_m},\ \forall k,\forall m \tag{21}$$

根据算力节点、转发时刻、路由路径的调度决策，任务 $k$ 的总完成时延和完成时刻为：

$$D_k^{\mathrm{sum}}=D_k^{\mathrm{fw}}+D_k^{\mathrm{tr}}+D_k^{\mathrm{al}}+\sum_{m\in\mathcal{M}}\sum_{t\in\mathcal{T}}\delta_{k,m,t}D_{k,m,t}^{\mathrm{q}}+\sum_{m\in\mathcal{M}}x_{k,m}D_{k,m}^{\mathrm{c}},\ \forall k \tag{22}$$

$$T_k^{\mathrm{f}}=T_k^{\mathrm{arr}}+D_k^{\mathrm{sum}},\ \forall k \tag{23}$$

式中，$D_k^{\mathrm{sum}}$ 是算力任务 $k$ 的总完成时延；$T_k^{\mathrm{f}}$ 是算力任务 $k$ 的完成时刻；$\mathcal{M}$ 为算力节点集合。

为完成算力任务的服务质量要求，完成时刻不得超过任务截止时间：

$$T_k^{\mathrm{f}}\le\overline{T}_k\ \ \forall k \tag{24}$$

2.2.6 算力网络能耗与供能模型

算力网络能耗主要由任务网络传输和计算处理环节产生。

算力网络采用基于光纤链路的 IP 路由架构，IP 路由器在电域进行数据交换与转发，数据在光纤链路中以光信号形式传输，因此任务数据在算力网络路由器需进行电子处理，还需通过应答器完成光电/电光信号转换，产生电子处理能耗和信号转换能耗，如式(25)-(26)所示：

$$P_{k,p,t}^{\mathrm{tr}}=\frac{d_k}{\Delta t}\left[\phi_i^{\mathrm{IP}}+\sum_{j\in\mathcal{R}_p}\phi_j^{\mathrm{IP}}+2\left(\phi_i^{\mathrm{op}}+\phi_m^{\mathrm{op}}+\sum_{j\in\mathcal{R}_p}\phi_j^{\mathrm{op}}\right)\right],\ \forall k,\forall p,\forall t \tag{25}$$

$$P_k^{\mathrm{tr}}=\sum_{t\in\mathcal{T}}\sum_{p\in\mathcal{P}_k}y_{k,p}z_{k,t}P_{k,p,t}^{\mathrm{tr}}\,,\ \forall k \tag{26}$$

式中，$P_{k,p,t}^{\mathrm{tr}}$ 为任务 $k$ 在时段 $t$ 经过路径 $p$ 的传输功率；$P_k^{\mathrm{tr}}$ 为任务 $k$ 在实际调度方案下的网络传输功

率；$\mathcal{R}_p$ 为路径 $p$ 所经过的普通路由器集合；$\phi_i^{\text{IP}}$ 和 $\phi_j^{\text{IP}}$ 分别为边缘路由器 $i$ 和普通路由器 $j$ 对单位数据进行电子处理的能耗；$\phi_i^{\text{op}}$、$\phi_j^{\text{op}}$、$\phi_m^{\text{op}}$ 分别为边缘路由器 $i$、普通路由器 $j$、算力路由器 $m$ 中应答器对单位数据进行光电/电光转换的能耗。

算力节点在执行计算任务过程中产生计算能耗，其大小与节点计算速率和完成的计算量有关。根据处理器动态功耗特性，算力节点 $m$ 在时段 $t$ 的计算功率为：

$$P_{m,t}^{\text{IT}}=\frac{\zeta_m f_m^2 W_{m,t}}{\Delta t},\ \forall m,\forall t \tag{27}$$

式中，$P_{m,t}^{\text{IT}}$ 为算力节点 $m$ 在时段 $t$ 的计算功率；$\zeta_m$ 为算力节点 $m$ 的计算芯片等效电容参数。

算力节点由本地的可再生能源和电力网络共同供电，如式(28)-(30)所示：

$$P_{m,t}^{\text{out}}=P_{m,t}^{\text{IT}}-P_{m,t}^{\text{R}},\ \forall m,\forall t \tag{28}$$

$$0\le P_{m,t}^{\text{R}}\le \overline{P}_{m,t}^{\text{R}},\ \forall m,\forall t \tag{29}$$

$$0\le P_{m,t}^{\text{out}},\ \forall m,\forall t \tag{30}$$

式中，$P_{m,t}^{\text{out}}$ 为算力节点 $m$ 在时段 $t$ 电力网络供电功率；$P_{m,t}^{\text{R}}$ 为算力节点 $m$ 在时段 $t$ 实际就地消纳的可再生能源功率；$\overline{P}_{m,t}^{\text{R}}$ 为算力节点 $m$ 所在区域在时段 $t$ 的可再生能源出力上限。

### 2.3 电力网络模型

本小节建立电力网络的数学模型[22]。

燃煤机组输出功率约束为：

$$0\le P_{g,t}^{\text{G}}\le P_g^{\max},\ \forall g,\forall t \tag{31}$$

式中：$P_{g,t}^{\text{G}}$ 为燃煤机组 $g$ 在时段 $t$ 的出力；$P_g^{\max}$ 为燃煤机组 $g$ 出力最大值。

输电线路容量约束为：

$$-F_l^{\max}\le P_{l,t}^{\text{L}}\le F_l^{\max},\ \forall l,\forall t \tag{32}$$

式中：$P_{l,t}^{\text{L}}$ 为输电线路 $l$ 在时段 $t$ 的传输功率；$F_l^{\max}$ 为输电线路 $l$ 的最大传输功率。

输电线路的直流潮流约束为：

$$P_{l,t}^{\text{L}}\cdot X_l=S_{\text{base}}\left(\theta_{l_{\text{from}},t}-\theta_{l_{\text{to}},t}\right),\ \forall l,\forall t \tag{33}$$

式中：$X_l$ 为线路 $l$ 的电抗；$S_{\text{base}}$ 为系统基准功率；$\theta_{l_{\text{from}},t}$ 和 $\theta_{l_{\text{to}},t}$ 分别为线路 $l$ 首、末节点在时段 $t$ 的电压相角。

节点功率平衡约束为：

$$\sum_{l\in\mathcal{L}}K_{b,l}^{\text{L}}P_{l,t}^{\text{L}}=\sum_{g\in\mathcal{G}}K_{b,g}^{\text{P}}P_{g,t}^{\text{G}}-P_{b,t}^{\text{D}},\ \forall b,\forall t \tag{34}$$

式中：$\mathcal{L}$ 为输电线路组合；$\mathcal{G}$ 为燃煤机组组合；$P_{b,t}^{\text{D}}$ 为电力节点 $b$ 在时段 $t$ 的负荷功率；$K_{b,l}^{\text{L}}$、$K_{b,g}^{\text{P}}$ 分别为电力节点-输电线路和电力节点-燃煤机组关联矩阵。

节点电压相角约束为：

$$-\pi\le\theta_{b,t}\le\pi,\ \forall b,\forall t \tag{35}$$

$$\theta_{b_{\text{ref}},t}=0,\ \forall t \tag{36}$$

式中：$\theta_{b,t}$ 为电力节点 $b$ 在时段 $t$ 的电压相角；$\theta_{b_{\text{ref}},t}$ 为电力参考节点在时段 $t$ 的电压相角。

燃煤机组运行成本采用二次函数表示：

$$C_{g,t}=\alpha_{0,g}+\alpha_{1,g}P_{g,t}^{\text{G}}+\alpha_{2,g}\left(P_{g,t}^{\text{G}}\right)^2,\ \forall g,\forall t \tag{37}$$

式中：$C_{g,t}$ 为燃煤机组 $g$ 在时段 $t$ 的运行费用；$\alpha_{0,g}$、$\alpha_{1,g}$、$\alpha_{2,g}$ 分别为燃煤机组 $g$ 运行成本函数的系数。

### 2.4 耦合约束

算力网络和电力网络通过算力节点的用电负荷进行耦合，如式(38)所示。

$$P_{b_m,t}^{D}=P_{m,t}^{\text{out}},\ \forall m,\forall t \tag{38}$$

式中，$b_m$ 为算力节点 $m$ 所连接的电力节点。

## 3 算例分析

所提方法在两个测试系统进行了仿真验证：（1）6-7 节点电算协同系统和（2）118-47 节点电算协同系统。算例基于 Matlab R2024a 平台进行计算，计算机配置为 Intel Core i9-14900HX CPU（2.20 GHz）和 32 GB 内存，求解器是 Gurobi 13.0.2。

### 3.1 6-7 节点电算协同系统

6-7 节点电算协同系统拓扑图如图 2 所示。其中，算力网络有 1 个边缘路由器，2 个普通路由器，2 个算力路由器和 2 个算力节点；电力网络有 6 个电力节点，11 条输电线路，3 台燃煤发电机组和 1 台光伏发电机。1 台光伏发电机直接接入算力节点 S1，所发电量就地消纳。算力节点 S1 和 S2 的计算能力分别设定为 1800 Gcycle/s、3600 Gcycle/s，链路 A-B、A-C、B-C 的数据传输速度随时间变化。

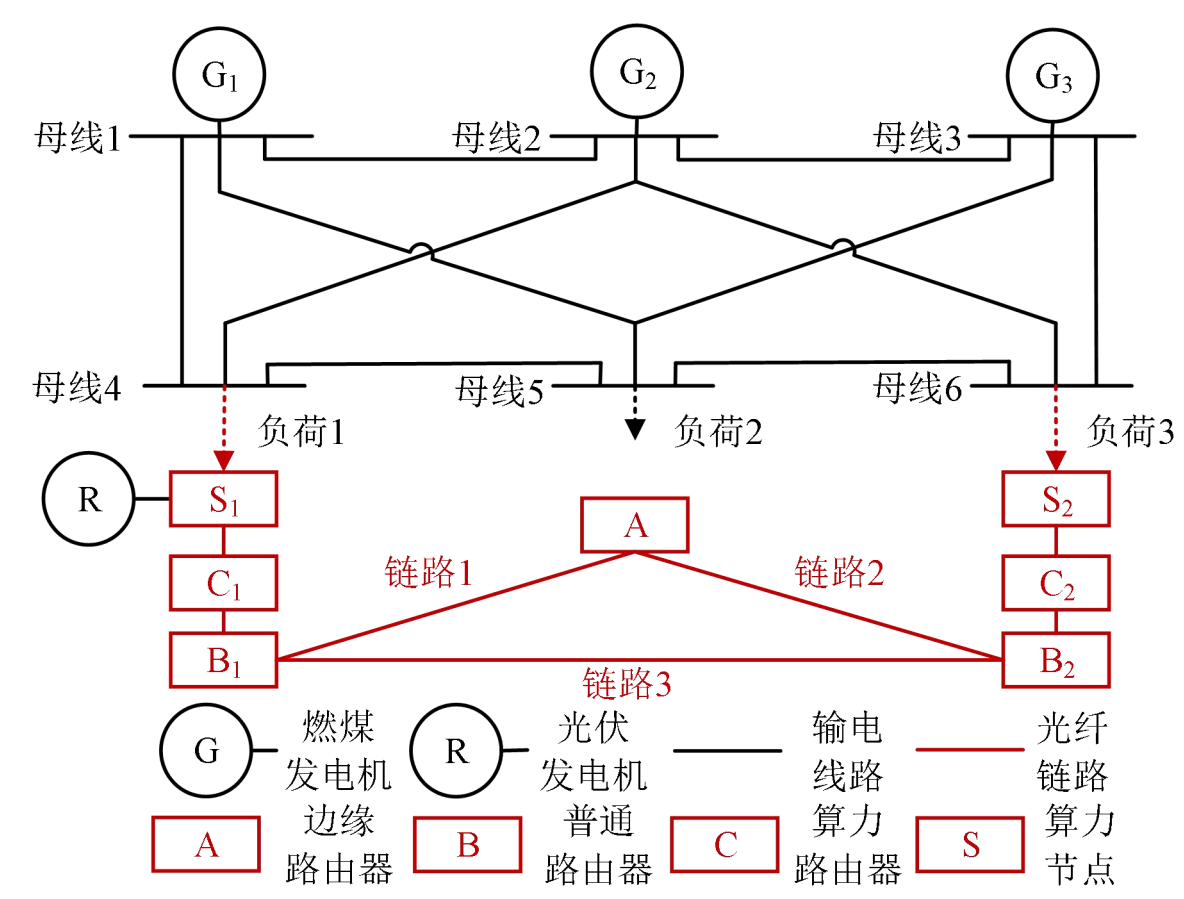

**图 2 6-7 节点电算协同系统拓扑图**

**Fig. 2 Topology of the 6-7 node power-computing coordinated system**

**表 1 算力任务基本参数**

**Tab. 1 Basic parameters of computing tasks**

| 任务 | 到达时刻 | 截止时刻 | 数据量/GB | 计算量/Pcycle |
|---|---|---|---|---|
| 1 | 0：00 | 4：00 | 20.00 | 32.40 |
| 2 | 1：00 | 17：00 | 35.00 | 36.00 |
| 3 | 6：00 | 12：00 | 22.00 | 9.60 |
| 4 | 12：00 | 19：00 | 26.00 | 19.20 |
| 5 | 1：00 | 7：00 | 25.00 | 38.40 |

为说明电算协同系统中算力任务的“传输-缓存-计算”流程，基于 6-7 节点电算协同系统，设置了 5 个具有不同数据规模和计算需求的算力任务，具体如表 1 所示。

**表 2 算力网络能耗及供能构成**

**Tab. 2 Energy consumption and supply composition of computing nodes**

| 算力节点 | 传输能耗/kWh | 计算能耗/kWh | 可再生能源供电/kWh | 燃煤发电机供电/kWh |
|---|---|---|---|---|
| S1 | 0.012 | 360.000 | 278.330 | 81.670 |
| S2 | 0.006 | 393.330 | 0 | 393.330 |
| 合计 | 0.018 | 753.330 | 278.330 | 475.000 |

表 2 展示了算力网络的能耗和供能构成。相较于算力节点的计算能耗，网络传输能耗的量级较小，对电力网络负荷的影响有限。因此，本文对网络传输能耗独立量化，不将其计入算力节点的本地供电平衡。算力节点 S1 总负荷为 360.000 kWh，由光伏和燃煤发电机共同供电，算力节点 S2 总负荷为 393.330 kWh，仅由燃煤发电机组供电。

图 3 展示了各个算力节点在不同时段的供电功率，由图 3 可知，算力节点 S1 的负荷集中在 7：00-18：00，与光伏出力时段相吻合，主要用于处理任务 2、3、4。算力节点 S2 的负荷则分布于 1：00-7：00，主要用于处理任务 1 和任务 5。在能耗表现上，由于算力节点 S2 具有更强的计算能力，在任务集中处理期间表现出更高的供电功率，峰值功率为 72 kW，显著高于算力节点 S1 36 kW 的峰值功率。

图 4 给出了两个算力节点的任务到达与计算处理过程。任务 3、任务 2、任务 4 经边缘路由器 A 转发，分别于 7：00、9：00、14：00 陆续到达算力节点 S1，进入计算队列等待处理。任务 1、任务 5 经转发分别于 1：00、2：00 到达算力节点 S2，排队后依次处理。算力节点 S2 的每小时的最大处理计算量为 12.96 Pcycle，是 S1 算力节点 6.48 Pcycle 的两倍，能够以更高的处理速率完成计算任务，更适合承担时间裕度较小的任务。

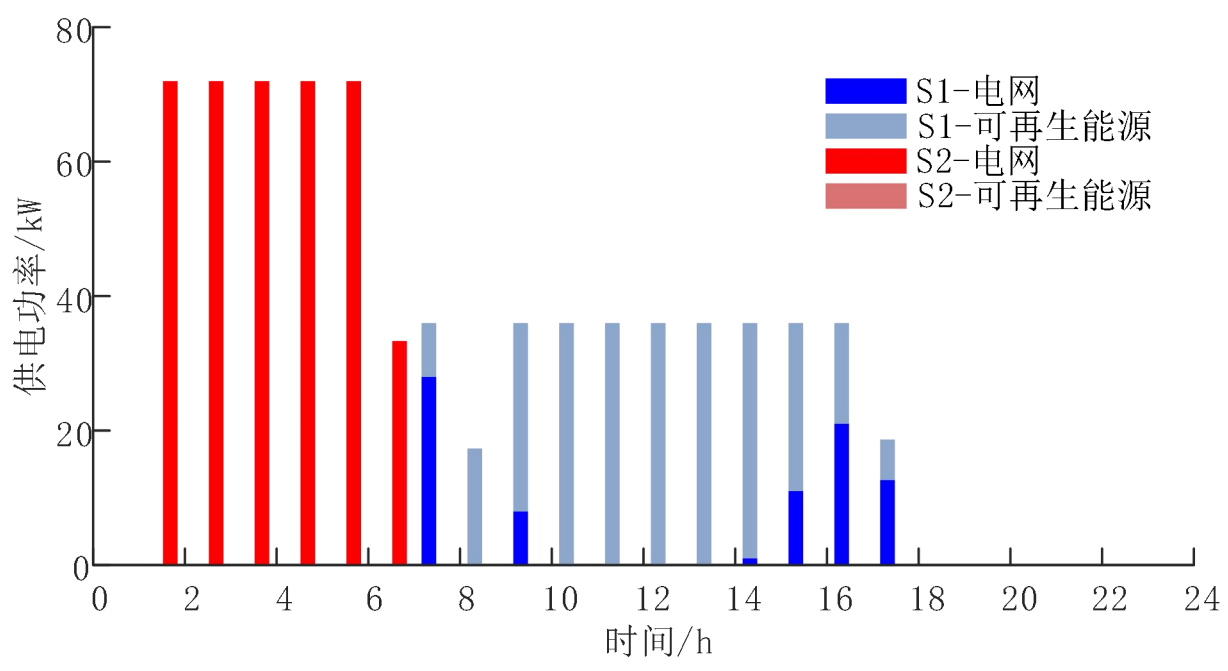

**图 3 各个算力节点各时段供电功率**

**Fig. 3 Hourly power supply of computing nodes**

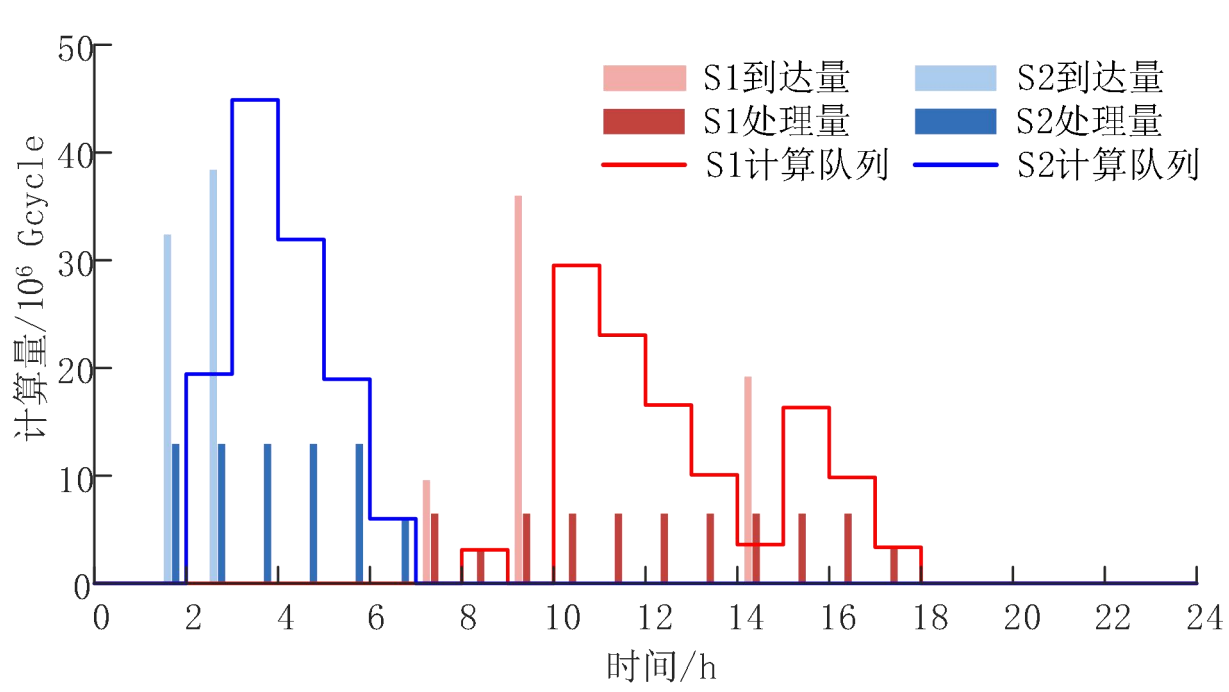

**图 4 算力节点任务到达与计算处理过程**

**Fig. 4 Task arrivals and computational processing at computing nodes**

**表 3 算力任务的时空调度结果**

**Tab. 3 Spatiotemporal Dispatch Results of Tasks in the Computing Network**

| 任务 | 算力节点 | 路径 | 转发时刻 |
|---|---|---|---|
| 1 | S2 | A-C-E-S2 | 0：00 |
| 2 | S1 | A-B-D-S1 | 8：00 |
| 3 | S1 | A-C-B-D-S1 | 6：00 |
| 4 | S1 | A-B-D-S1 | 13：00 |
| 5 | S2 | A-C-E-S2 | 1：00 |

表 3 给出了各算力任务对应的算力节点、路由路径和转发时刻。任务 1 和任务 5 的截止时刻较近，时延裕度较小，到达后被立刻转发到计算能力较强的算力节点 S2 处理，以保证在截止时刻前完成。任务 2、任务 3 和任务 4 的截止时刻较为宽松，被调度至与光伏发电机直接相连的算力节点 S1，通过合理调整任务的转发时刻，使计算负荷尽可能与光伏出力时段相匹配，从而充分消纳光伏发电。

在路由路径选择方面，任务 1、2、4、5 均选择了所含链路数较少的路径，任务 3 选择了路径 A-C-B-D-S1。对于任务 3，存在 A-B-D-S1 和 A-C-B-D-S1 两条候选路径，由于在 6：00-7：00 时段内链路 A-B、A-C、C-B 传输速率分别为 40、145、130 Mbit/s，受链路时变传输能力的影响，虽然路径 A-C-B-D-S1 链路更多，但其总网络传输时延反而更短，所以任务 3 选择该路径进行转发。

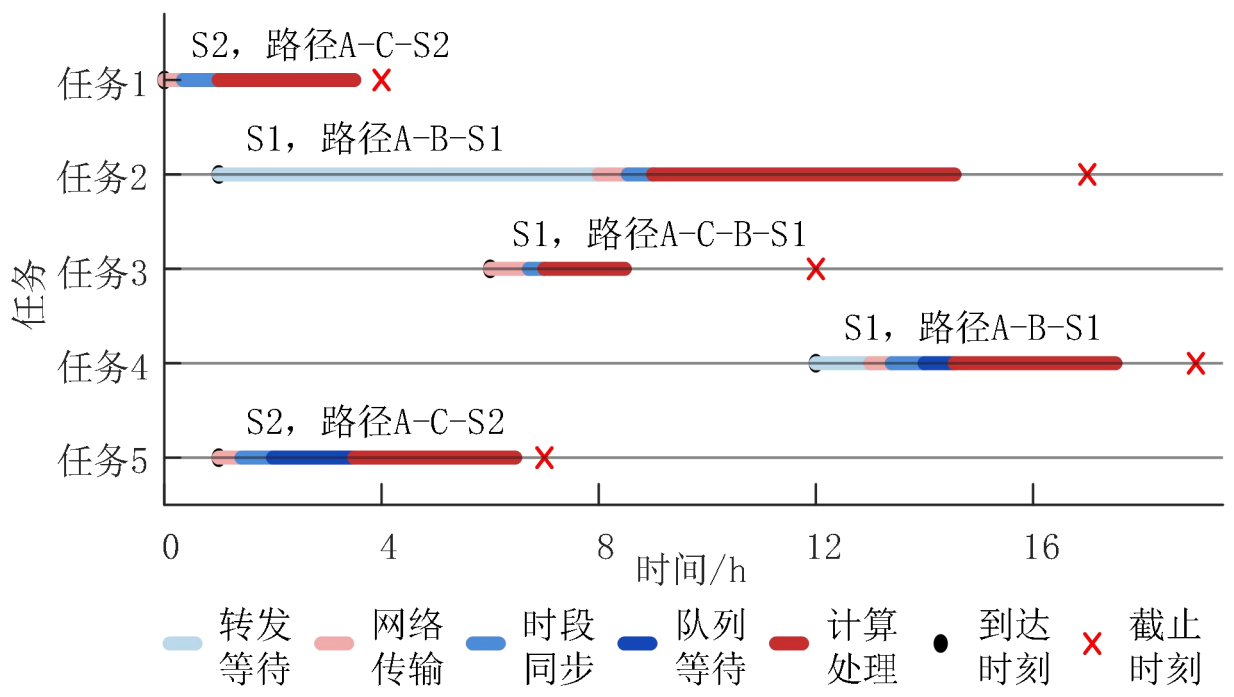

**图 5 算力任务全流程处理时序**

**Fig. 5 Overall processing timeline of computing tasks**

图 5 展示了各算力任务从到达边缘节点开始，依次经历传输、等待和计算直至完成的全时序过程。如图 5 所示，任务 2 在 1：00 就到达了边缘路由器 A，截止时刻在 17：00，时延裕度较大，因此为充分利用光伏发电，该任务在 8：00 时才进行转发，经历 0.54 h 的网络传输时延到达算力节点 S1，此时算力节点 S1 计算队列为 0，任务无需排队等待，计算处理时延为 5.55 h，最终于 14：33 时完成处理。任务 5 在 1：00 到达边缘节点 A，计算量为 38.40 Pcycle，要在 6 h 内处理完成，时间裕度较小。该任务在算力节点 S1、S2 满功率处理分别需要 5.93 h 和 2.96 h，所以被立刻转发，经历 0.42 h 的网络传输时延到达计算能力更强的算力节点 S2，此时算力节点 S2 还未处理完任务 1，任务排队等待 1.50 h 后开始处理，计算处理时延为 2.96 h，最后在 6：28 处理结束。

### 3.2 118-47 节点电算协同系统

算力网络有 5 个边缘路由器，22 个普通路由器，10 个算力路由器，10 个算力节点，拓扑基于欧洲的 GÉANT 网络进行构建[23]；电力网络采用 IEEE 118 节点系统，有 118 个电力节点，186 条输电线路，54 台燃煤发电机组，4 台光伏发电机组和 3 台风力发电机组。各个算力节点所处地理位置不同，可再生资源禀赋存在差异，可再生能源机组据此接入相应的算力节点：算力节点 1、4、7、10 所在区域具有较丰富的太阳能资源，配置光伏发电机组，算力节点 2、5、8 所在区域具有较丰富的风能资源，配置风力发电机组。

为了验证所提出的电算协同优化方法的有效性，基于 118-47 节点电算协同系统，设置了 4 个场景，具体为

场景I：基础场景，不考虑算力任务的时空灵活性，算力任务按照初始调度方案执行。

场景II：仅考虑算力任务的时间灵活性，优化算力任务的转发时间。

场景III：仅考虑算力任务的空间灵活性，优化算力任务的算力节点、路由路径。

场景IV：考虑算力任务的时空灵活性，优化算力任务的转发时间、算力节点、路由路径。

**表 4 不同场景下系统运行性能对比**

**Tab. 4 Comparison of system operational performance under different scenarios**

| 场景 | 可再生能源供能/MWh | 电网供能/MWh | 可再生能源消纳率/% | 系统供电成本/元 | 降低率/% |
|---|---|---|---|---|---|
| I | 206.70 | 1321.88 | 21.19 | 55145.59 | 0 |
| II | 414.41 | 1114.16 | 42.49 | 50846.25 | 7.80 |
| III | 827.95 | 700.62 | 84.90 | 42379.68 | 23.15 |
| IV | 945.10 | 583.47 | 96.91 | 39984.21 | 27.49 |

| 场景 | 转发等待时延/h | 网络传输时延/h | 时段同步时延/h | 队列等待时延/h | 计算处理时延/h | 任务平均完成时延/h |
|---|---|---|---|---|---|---|
| I | 0.00 | 0.28 | 0.72 | 0.67 | 0.17 | 1.84 |
| II | 1.03 | 0.28 | 0.72 | 0.38 | 0.17 | 2.58 |
| III | 0.00 | 0.45 | 0.57 | 1.75 | 0.20 | 2.97 |
| IV | 1.23 | 0.42 | 0.58 | 0.06 | 0.18 | 2.47 |

表 4 对比了不同场景下系统运行性能。在场景I中，算力任务到达后选择最短的传输路径立即转发到距离最近的算力节点进行处理。由于缺乏电力网络和算力网络的协同优化，算力任务无法主动匹配可再生能源的出力变化，导致系统总成本最高，为 55145.59 元，可再生能源消纳率最低，为 21.19%。

在场景II中，仅考虑算力任务的时间迁移能力，在保持算力节点和传输路径不变的情况下，通过调整任务转发时间以匹配可再生能源出力。相比场景I，能利用可再生能源高出力时段执行算力任务，提升可再生能源消纳水平，使系统总成本降低了 7.80%，可再生 n 能源消纳率提升至

42.49%。然而由于任务只能在时间维度进行调整，部分任务需要推迟执行，导致转发等待时延增加至 1.03 h，任务平均完成时延增加至 2.58 h。

在场景III中，仅考虑算力任务的空间迁移能力，即在任务到达后立即转发，但可根据不同算力节点的可再生能源出力情况优化算力节点和传输路径。该场景可以将部分算力任务迁移至风光资源较丰富的算力节点，进一步消纳可再生能源。相比场景I，系统总成本降低了 23.15%，可再生消纳率提升至 84.90%。然而由于缺少时间维度的调节能力，部分新能源富集区域的算力节点承担了大量任务，造成任务处理队列拥塞，使队列等待时延达到 1.75 h，为 4 个场景中的最大值，任务平均完成时延增加至 2.97 h。

在场景IV中，考虑算力任务的时空迁移能力，联合优化算力节点、传输路径和转发时刻，实现算力网络与电力网络的协同优化。该场景能够在新能源高出力时段执行算力任务，并根据不同区域可再生能源出力情况调整任务的空间分布，避免算力节点处理队列拥塞。相比场景I，系统总成本降低了 27.49%，可再生能源消纳率提升至 96.91%。通过时间维度上的主动调节，部分任务被推迟转发，转发等待时延增加至 1.23 h，为 4 个场景中的最大值，不过由于空间迁移优化了任务在各算力节点间的分配，队列等待时延降低至为 0.06 h，任务平均完成时延为 2.47 h。

综上所述，所提电算协同优化方法可以通过联合优化任务转发时刻、路由路径和目标算力节点，在提升可再生能源消纳能力的同时，有效协调计算资源分配，实现电力网络供电经济性与算力网络服务质量的协同优化。

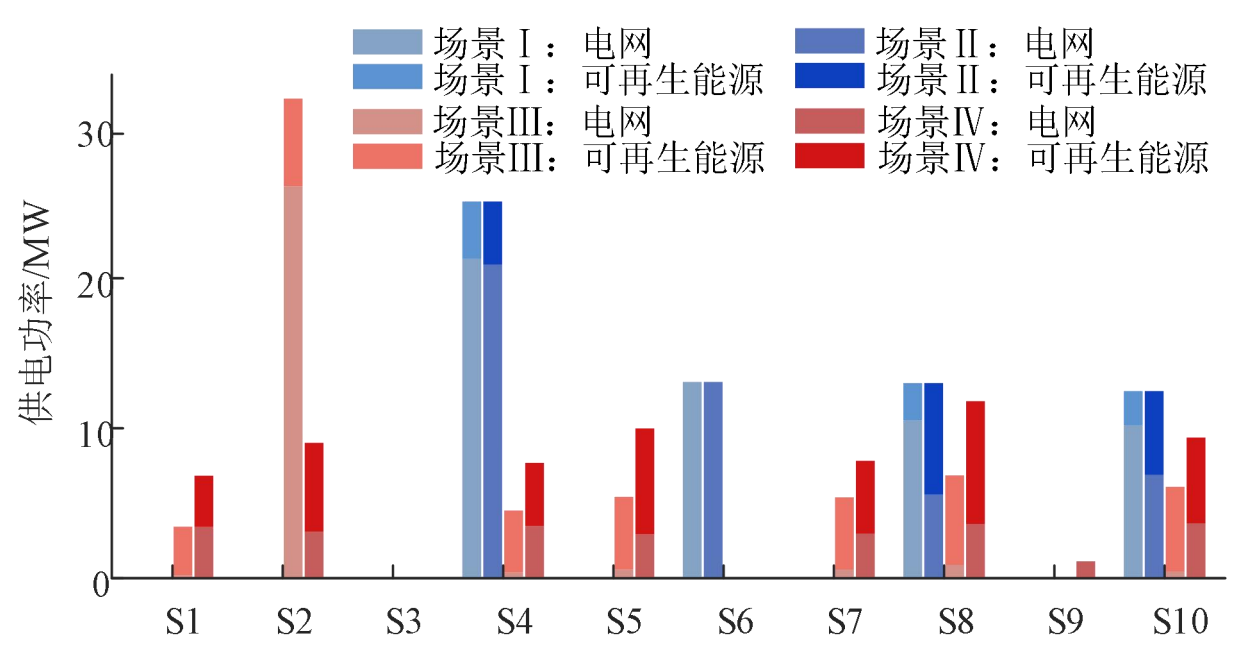

**图 6 不同场景下各个算力节点供电功率**

**Fig. 6 Hourly power supply of computing nodes under different scenarios**

图 6 对比了不同场景下各个算力节点的供电功率。在场景I和场景II中，算力任务通过边缘路由器转发到最近的算力节点 S4、S6、S8、S10 来进行处理，各个算力节点的供电功率相同，但是场景II中算力任务可以优化转发时间匹配可再生能源出力，从而提高可再生能源供电比例。在场景III和场景IV中，算力任务可以空间迁移，选择合适的路由路径和算力节点，算力任务更多转发至配置风电或光伏的算力节点，新能源供电比例进一步提升。

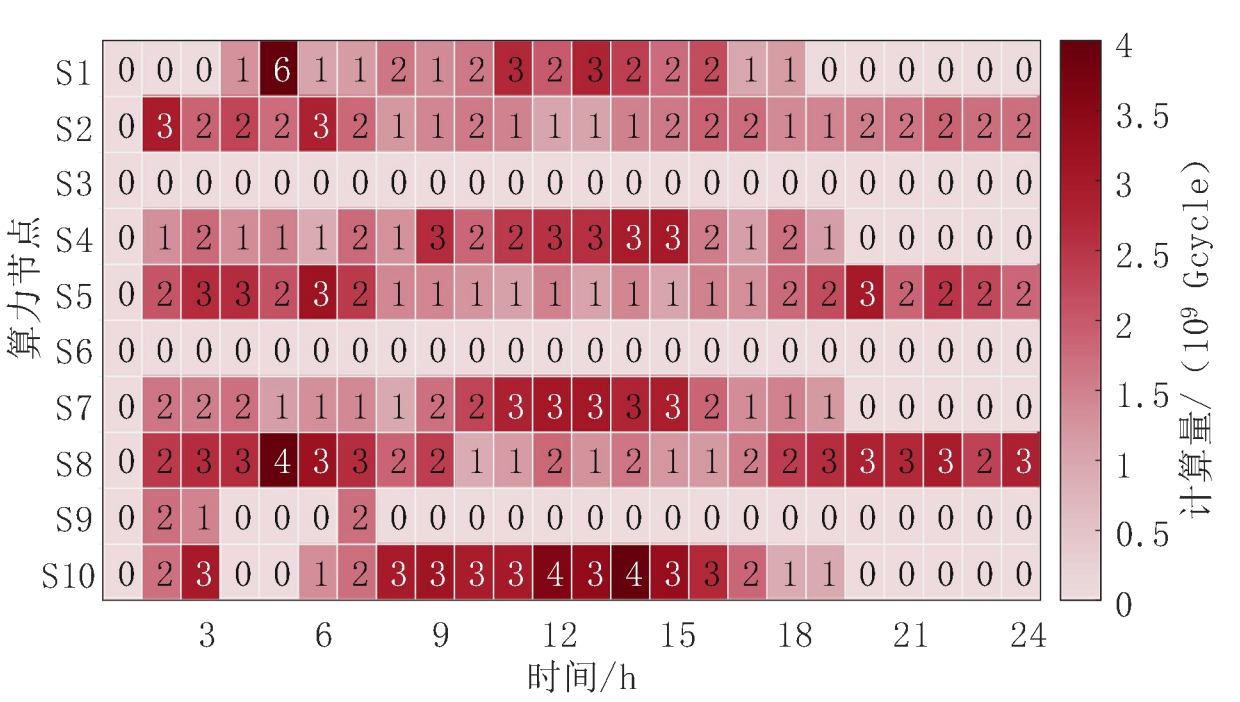

**图 7 场景IV算力节点各时段算力处理量**

**Fig. 7 Hourly processed computing workload of computing nodes in Scenario IV**

图 7 展示了场景IV中各个算力节点各时段的算力处理量。算力节点 S1、S4、S7 和 S10，有着丰富的光伏资源，并与光伏机组直接相连，算力任务大多在光伏出力时间 6：00-19：00 被处理。算力节点 S2、S5 和 S8，有着丰富的风力资源，并与风电机组直接相连，算力任务大多在风电出力 1：00-8：00、18：00-24：00 被处理。仅由电网供电的算力节点 S9，仍然承担着少量的算力任务，避免其余算力节点任务过度集中，从而减少队列等待时延，保证算力任务尽快处理。

## 4 结论

本文提出一种计及算力任务端到端完成时延的电算协同优化运行方法，引导算力负荷与可再生能源在时空上协调匹配。首先，统一描述算力任务“传输-缓存-计算”全流程，准确量化端到端完成时延。其次，联合优化任务转发时段、路由路径、目标算力节点，实现算力负荷任务级的时空调节。然后，兼顾电力网络运行经济性与算力网络的服务时效性，构建了电算协同优化模型。算例结果证明了所提出方法的有效性，能够充分挖掘算力负荷任务级时空调度潜力，在降低系统供电成本的同时保证算力任务服务质量。未来的研究将进一步考虑算力任务到达及可再生能源出力的不确定性，提升调度方案的鲁棒性。

## 参考文献